\documentclass{article}
\font\cero=cmss10 scaled 1728 
\usepackage{tensor}
\usepackage{verbatim}
\usepackage{amsfonts}
\usepackage{graphicx}
\usepackage{amssymb}
\usepackage{float}
\begin{document}
\begin{flushleft}
{\cero Hyperbolic deformation of a gauge field theory and the hierarchy problem}\\
\end{flushleft}
{\sf R. Cartas-Fuentevilla, A. Escalante-Hernandez, and A. Herrera-Aguilar}\\
{\it Instituto de F\'{\i}sica, Universidad Aut\'onoma de Puebla,
Apartado postal J-48 72570 Puebla Pue., M\'exico}; 

E-mail: rcartas@ifuap.buap.mx

ABSTRACT: The problem of the gauge hierarchy is brought up in a  hypercomplex scheme for a $U(1)$ field theory; in such a scheme a compact gauge group is deformed through a $\gamma$-parameter that varies along a non-compact internal direction, transverse to the $U(1)$ compact one, and thus an additional $SO(1,1)$ gauge symmetry is incorporated. This transverse direction  can be understood as an  {\it extra internal dimension}, which will control the spontaneous symmetry breakdown, and will allow us to establish a mass hierarchy. 
In this mechanism there is no brane separation to be estabilized as in the braneworld paradigm, however, a different kind of fine-tuning is needed in order to generate the wished electroweak/Planck hierarchy. By analyzing the effective self-interactions and mass terms of the theory, an interesting duality is revealed between the real and hybrid parts of the effective potential.
This duality relates the weak and strong self-interaction regimes of the theory,
due to the fact that both mass terms and self-coupling constants appear as one-parameter flows in $\gamma$.
Additionally the $\gamma$-deformation will establish a flow for the electromagnetic coupling that mimics the renormalization group flow for the charge in QED.\\


\section {Introduction}
\label{intro}
It is well known that the physics of fields that arise from the dimensional reduction of M/string theory depends sensitively on the geometry and topology of the underlying space; in the context of Calabi-Yau compactification, the emerging theories contain supersymmetric multiplets and hypermultiplets, which can be described by defining a symplectic vectorial space where those fields appear as vectors and scalars; the inner product on such a space has the form of the modulus of split-complex numbers, rather than of
the ordinary complex numbers \cite{emam1,emam2,emam3,emam4,emam5}. The split-numbers can be generated by the substitution $i\rightarrow j$, where $j$ is a new complex unit with the property $j^2=1$; hence, one obtains an appropriate formulation for generating solutions and to gain understanding of the general structure of the theory, {\it i.e.} of the string/M theory landscape. Similarly the appropriate formulation
of the supergravity description of D-instantons  in the context of type IIB superstring theory, requires the same mysterious substitution rule in the Lagrangian and in the supersymmetric rules \cite{perry}. Furthermore, 
a generalization of the ring of the split-complex numbers can be generated by the incorporation of the new complex unit, instead of the simple substitution mentioned earlier; the ring generalized for hypercomplex numbers contain thus two complex units. This extended complexification allows a new description of fermions and bosons, with the possibility of new interactions arising from hypercomplex gauge transformations \cite{ulrych}. From the mathematical point of view there exist formally three different complex units, namely, the (conventional) elliptic $i^2=-1$, the hyperbolic $j^2=1$, and the parabolic, for which the square of the complex unit vanishes (see for example, \cite{kisil}).

With these antecedents, 
 we have recently developed a hypercomplex formulation of Abelian gauge field theories, by incorporating the new complex unit to the usual complex one \cite{1}. Physically the hypercomplex formulation allows us to accommodate hyperbolic complex counterparts for the usual $U(1)$ interactions, and hence to explore their possible realizations beyond presently known energies. In the particular case of the hypercomplex electrodynamics, the results show exotic scenarios for spontaneous symmetry breaking, such as running masses for vectorial and scalar fields, that mimic flows of the renormalization group. Additionally the so called Aharonov-Bohm 
string defects emerge as possible topological defects admitted in the new formulation; such defects are detectable in principle only by quantum interference.
By using the commutative ring of hypercomplex numbers, the usual real objects such as Lagrangians,  
vector fields, the norm of a complex field, masses, coupling parameters, etc, are generalized to {\it Hermitian} objects, encoding two real quantities.
Furthermore, in this scheme, a hypercomplex field will have four real components, instead of the two real components of an usual complex field;
however, in the hypercomplex formulation developed in \cite{1}, those four components are identified to each other using a real dimensionless $\gamma$-parameter, leading to two real effective variables. Therefore, the new formulation is constructed as a $\gamma$-deformation of the $U(1)$- formulation of an Abelian gauge theory; the deformation implies the incorporation of a new symmetry, namely, the hyperbolic rotations as a complement of the  circular $U(1)$-rotations; the full symmetry group will correspond at the end to $SO(1,1)\times U(1)$, the product of a noncompact group and a compact one. The effect of such a $\gamma$-deformation is visible directly on the profile of the familiar $U(1)$ hat potential, which is {\it hallowed out} at two points in the valley that defines the degenerate vacuum; the new vacuum is moved on such points; the depth of the new vacuum, the vacuum expectation values of the fields, and the masses obtained by spontaneous symmetry breakdown, are determined by specific polynomials in $\gamma$.

 \section{ Motivations and an advance of results}
\label{moti}

There are two general frameworks for facing the mass hierarchy problem, namely, invoking new (super-)symmetries at the high-energy regime \cite{wittensusy}, and involving extra-dimensional spaces with either flat \cite{anto,nima,antonima} or curved backgrounds \cite{rs}. 

Within the first approach the large ratio between the mass scale of particle physics, which is assumed to be fundamental, and the Planck scale is dynamically generated as long as supersymmetry is spontaneously broken at the tree level.\footnote{Namely, a natural gauge hierarchy, but unfortunately not the right one, is obtained when SU(5) is strongly broken to SU(3)XU(1)XU(1) which in turn is weakly broken to SU(3)XU(1). In this picture, the Higgs boson is light, while its SU(5) partners automatically get large masses through the vev of a complex field.}

Within the paradigm of models which make use of a number of flat extra dimensions, gauge and gravitational interactions get unified at the weak energy scale, which is considered as the only fundamental energy scale, and gravity is effectively weak at distances greater than a millimeter due to the large scale character of the extra dimensions compared to the weak scale; moreover, in this picture the dynamics of the Standard Model fields is localized on a 4D hypersurface with weak scale interactions, while gravity, which describes the dynamics of the spacetime itself, can propagate in the whole bulk space. Nevertheless, in this flat extra dimensional paradigm, one needs at least three higher dimensions for matching the LHC experimental data.

If the extra dimensions are warped, the need for certain number of extra higher dimensions is reduced to one, requiring as well a bulk cosmological constant and a pair of gravitationally coupled delta function thin branes (4D hypersdurfaces which can represent our Universe) \cite{rs}. Thus, the dynamics of the 4D Standard Model particles takes place in one of the aforementioned 4D branes, representing our world. However, in this model, the 5D manifold is an orbifold and, hence, possesses naked singularities at its singular points, i.e. at the brane positions where the branes with opposite tensions are placed. Several models that make use of scalar fields have emerged in the literature with the aim of solving this problem by smoothing out the braneworld configurations since from the gravitational point of view, we can not live in a singularity. Moreover, thick braneworld models with two positive branes have solved the mass hierarchy problem within this context \cite{bhnkq}. In this picture, the hierarchy between the Planck and the Tev energy scales emerges by placing a 4D TeV brane some distance away 
from the Planck brane, where gravity is localized. It should mentioned as well that within this setup the introduction of the TeV brane away from the Planck one 
gives rise to the so-called brane separation stability problem, consisting of a new fine-tuning on the TeV brane position and, therefore, to the need the need of stabilizing this brane separation. This brane separation remains stable when modeled by a scalar field with a family of quartic self-interacting potentials \cite{dewolfe}. 

The proposal in this paper is close in spirit to the former paradigm: we construct a simple model in which the hyperbolic rotations correspond to a non-compact new symmetry, whose inclusion will deform the low-energy $U(1)$ gauge symmetry; the mass hierarchy will be originated then from an unusual spontaneous symmetry breakdown of the new symmetry. 

It is worth mentioning that within the framework of this hyperbolic definition of gauge field theories, there is no need to stabilize any brane separation; however, another kind of fine-tuning is needed in order to generate the electroweak/Planck hierarchy.

Although the above mentioned $\gamma$-parameter plays a crucial role in regulating the deformation of the theory,  it played a passive role in the pattern of the spontaneous symmetry breakdown obtained in \cite{1}, since the diagonalization of the mass matrix was realized by choosing a gauge that involves
only the circular and hyperbolic parameters of the symmetry group $U(1)\times SO(1,1)$, according to the conventional procedure for such a purpose. However, as we shall see in the present work, it is possible to generate a qualitatively different pattern of SSB controlled by the same $\gamma$-parameter, introducing naturally a hierarchy of orders of magnitude in the mass spectrum.  The $\gamma$-parameter varies smoothly within a range determined in \cite{1} in order to induce a stable vacuum; the determination of the hierarchy  is then transformed into the choice of the adjustable $\gamma$-parameter. All physical quantities turn out to be effective functions of the $\gamma$-parameter; hence, we shall find multiple criteria, for example,  imposing masslessness for certain fields, local maxima and minima for the effective masses, effective couplings, vacuum expectation values, and for vacuum energies. 
 Thus, these results constitute the main motivation of the present work, and represent an important extension of the results obtained previously in \cite{1}. 

In section \ref{lambda4} we develop the hypercomplex formulation for 
a charged scalar field using the commutative ring for split complex numbers; this section is basically an outline of the formulation developed in \cite{1}, emphasizing the  aspects that are relevant for the present treatment. In this section
the $\gamma$-parameter appears deforming a $U(1)$ gauge field theory, in such a way that a non-compact gauge group is incorporated. The profile of the potentials  bounded from below, and the appearance of a new degenerate vacuum is outlined; the spontaneous symmetry breakdown of global symmetries by using a conventional gauge is described; in this gauge the $\gamma$-parameter plays no role. However the same scheme allows us to use an unusual gauge, which is controlled by the $\gamma$-parameter; the mass spectrum generated in each case will show physically different properties. These results are established in Sections \ref{gammano1}, and \ref{no1} for different restrictions on the {\it Hermitian} parameters that define the hypercomplex gauge field theory. In the respective sub-sections, the unusual gauge will allow us to generate a mass hierarchy, by fixing the value of the $\gamma$-parameter, and a pair of certain discrete parameters $(l=\pm 1,s= \pm 1)$. A way of solving the gauge hierarchy problem is outlined by fine-tuning the $\gamma$-parameter and assigning electroweak mass scales to one field and Planck scales to the other one. 
In Section \ref{nogamma11}  we briefly describe alternative cases with a different choice for the parameters that define the model. In the Section \ref{hed} the local circular and hyperbolic rotations are considered by coupling the split-complex
scalar field to vectorial boson fields. The mass hierarchy induced in the case of global symmetries can not be retained for the local case, but a hierarchy is originated by using a similar criterion. We finish in section \ref{cr} with some concluding remarks and open questions.

\section{ The model for a  hypercomplex field} 
\label{lambda4}
We describe briefly the hypercomplex formalism used in \cite{1}; the commutative ring of hypercomplex numbers is defined as
\begin{eqnarray}
     z =  x +  iy + jv + kw, \quad \overline{z}  =  x - iy -jv + k w, \quad  x,y,v,w \in {\cal R} ,
          \label{ring}
\end{eqnarray} where the hyperbolic unit $j$ has the properties $j^{2}=1$, and $\overline{j} = -j$, and, as usual, $i^{2}=-1$, and $\overline{i}=-i$;
the hybrid unit $k\equiv ij$, is {\it Hermitian} $\overline{k}=k$, with $k^2=i^2j^2=-1$; we have additionally that  $ik=i^2j=-j$, and $jk=ij^2=i$.
 Hence,  the square of the hypercomplex number is given by
\begin{equation}
     z \overline{z} = x^{2} + y^{2} -v^{2} - w^{2} +2k (xw-yv),
     \label{square}
\end{equation} 
which is not a real number, instead it is in general a {\it Hermitian} number, which is the generalization of a simple real number in the hypercomplex scheme.
The expression (\ref{square}) is invariant under the usual circular rotations $e^{i\theta}$, represented by the Lie group $U(1)$, and under {\it hyperbolic} rotations that can be represented by the connected component of the Lie group $SO(1,1)$ containing the group unit. A hyperbolic rotation is represented by the hyperbolic versor $e^{j\chi} \equiv \cosh\chi + j \sinh\chi$, with the split-complex conjugate $e^{-j\chi} = \cosh\chi - j\sinh\chi$, and with the operations $e^{j\chi} \cdot e^{j\chi'} = e^{j(\chi +\chi')}$, where $\chi$ is a real parameter.
 
The identification $x=\gamma w$ and $y=\gamma v$, with $\gamma$ a real parameter,  reduces the expression (\ref{ring}) to
\begin{equation}
     z= (\gamma +k)w+(i\gamma +j)v , \quad \overline{z} = (\gamma +k)w-(i\gamma +j)v,\quad  z\overline{z} = (\gamma^{2}-1)(v^{2}+w^{2}) + 2k\gamma (w^{2}-v^{2});
     \label{correct}
\end{equation}
with only two effective variables; the norm is invariant under the interchange
of the field $v\leftrightarrow w$, and simultaneously the change $\gamma\rightarrow-\gamma$.

Therefore, the following Hermitian Lagrangian, invariant under global phases $e^{i\theta}e^{j\chi}$, where $\theta$ is also a real quantity, can be constructed for a hypercomplex field with
two dynamical variables,
\begin{eqnarray}
     {\cal L}(\psi,\overline{\psi}) = \int dx^{d} \Big[\frac{1}{2} \partial^{i} \psi \cdot \partial_{i}\overline{\psi} - V(\psi ,\overline{\psi})\Big], \qquad V(\psi ,\overline{\psi}) = \frac{a}{2} m^{2} \psi\overline{\psi} + \frac{\lambda}{4!} \psi^{2}\overline{\psi}^{2},
     \label{complexlag}\\
   \psi\overline{\psi}=(\gamma^2-1)(v^2+w^2)+2k\gamma(w^2-v^2),\label{thenorm}
\end{eqnarray}
the square mass and the coupling are {\it Hermitian}, with real and hybrid parts $m^2\equiv m^2_{R}+km^2_{H}$, $\lambda\equiv \lambda_{R}+k\lambda_{H}$, with $(m^2_{R}, m^2_{H}, \lambda_{R}, \lambda_{H}) $ real parameters;  $a=\pm 1$.
Due to the parameters and the norm of the fields (\ref{correct}) are valued in the sub-set of Hermitian numbers, the Lagrangian
(\ref{complexlag}) is valued in the same sub-set, and has the form $R+kR$. 

The symmetry under phase transformation leads to $U(1)\times SO(1,1)$-Noether (probability) current, which has the same functional 
dependence on the fields $(\psi, \bar{\psi})$ that appears in the usual  $U(1)$-Noether  current:\begin{equation}
     J_{\mu} \equiv \overline{\psi}\partial_{\mu}\psi - \partial_{\mu}\overline{\psi}\cdot\psi , \quad \partial_{\mu}J^{\mu} =0,
\label{noether}
\end{equation}

Although the formulation of non-real Lagrangian is no conventional,  the formulation of strict complex actions have been considered 
recently; for example, holomorphic models (in the conventional sense) can manifest a hidden gauge symmetry that connects different real systems \cite{vergara};  if the potential is holomorphic, then that symmetry is related to the Cauchy-Riemman
conditions; additionally a gauge condition determines the type of hermiticity  of the variables. In the present work we are not considering {\it holomorphic}  Lagrangian from the hypercomplex point of view, instead we are considering only the natural extension of the real Lagrangians in a hypercomplex scheme, namely Hermitian Lagrangians.  We shall see that the real components of the Hermitian quartic potential
in the Eq. (\ref{complexlag}) are connected by a duality, and a mass hierarchy is induced by spontaneous symmetry breakdown.

The potential can be written explicitly in terms of its real and hybrid parts as $V=V_{R}+kV_{H}$,
\begin{eqnarray}
     V _{R}  \!\! & = & \!\!  a\Big(\frac{\gamma^{2}-1}{2} m^{2}_{R} + \gamma m^{2}_{H}\Big)v^{2} + a\Big(\frac{\gamma^{2}-1}{2} m^{2}_{R} - \gamma m^{2}_{H}\Big) w^{2}\nonumber\\ 
               \!\! & + & \!\!   \frac{\lambda_{R}}{6} \Big[ \frac{(\gamma^{2}-1)^{2}}{4} (v^{2}+w^{2})^{2} - \gamma^{2} (v^{2}-w^{2})^{2}\Big] -\frac{\lambda_{H}}{6}
    \gamma(\gamma^2-1)(w^4-v^4); 
     \label{VR} \\
   V_{H}  \!\! & = & \!\!  a\Big(\frac{\gamma^{2}-1}{2} m^{2}_{H} -\gamma m^{2}_{R}\Big) v^{2} + a \Big(\frac{\gamma^{2}-1}{2} m^{2}_{H} + \gamma m^{2}_{R}\Big)w^{2} + \frac{\gamma\lambda_{R}}{6} (\gamma^{2}-1) (w^{4}-v^{4})\nonumber \\
   \!\! & +& \!\! \frac{\lambda_{H}}{6} \Big[ \frac{(\gamma^{2}-1)^{2}}{4} (v^{2}+w^{2})^{2} - \gamma^{2} (v^{2}-w^{2})^{2}\Big];
     \label{potentialVH}
\end{eqnarray}
one can map the potentials $V_{R}$ and $V_{H}$ to each other,  by the discrete transformations 
\begin{equation}
\gamma \rightarrow -\gamma,\quad 
(\lambda_{R},\lambda_{H})\rightarrow (\lambda_{H},\lambda_{R}), \quad (m_{R}, m_H)\rightarrow (m_{H},m_R).
\label{discrete}
\end{equation}  
At this point, it is not yet evident  that the potentials $V_{R}$ and $V_{H}$ are connected by a duality, which will be shown explicitly 
in its due course.

The vacuum is defined as usual by the stationary points constraint,
\begin{equation}
    \frac{\partial V}{\partial\psi_{0}} = \overline{\psi_{0}} [am^{2} + \frac{\lambda}{6} \psi_{0} \overline{\psi_{0}}] =0;
    \label{vacuum}
\end{equation}
therefore, the zero-energy point for $V_{R}$ and $V_{H}$ is described by
\begin{eqnarray}
      \Big( v_{0}=0,  w_{0}=0\Big); \label{zerozero}\\ 
       V_{R}=0;
      \quad ({\rm det} {\cal H})_{ (V_{R})}=4\Big[
     \big(\frac{\gamma^2-1}{2}\big)^2m^4_{R}-\gamma^2m^4_{H}\Big]; \label{falsev}\\
   \qquad V_{H} =0; \qquad \det {\cal H}_{ (V_{H})} = 4\Big[ \left(\frac{\gamma^{2}-1}{2}\right)^{2} m^{4}_{H} - \gamma^{2} m^{4}_{R}\Big]; 
\label{falsevv}      
\end{eqnarray} 
where we have displayed the determinant of the Hessian matrix $\cal H$, for each potential;
likewise,  the other stationary points for $V_{R}$ and $V_{H}$ related with the second condition in (\ref{vacuum}), $ am^{2} + \frac{\lambda}{6} \psi_{0} \overline{\psi_{0}}=0$, can be solved for the fields $(v_0,w_0)$:
\begin{eqnarray}
v_{0}^2=\frac{3}{2a} \frac{1}{\lambda_{R}^2+\lambda_{H}^2}\frac{1}{\gamma(1-\gamma^2)}\Big\{\Big[ (\gamma^2-1)\lambda_{H}+2\gamma\lambda_{R}             \Big]m_{R}^2+ \Big[ (1-\gamma^2)\lambda_{R}+2\gamma\lambda_{H}  \Big]m_{H}^2  \Big\},\nonumber \\
w_{0}^2=\frac{3}{2a} \frac{1}{\lambda_{R}^2+\lambda_{H}^2}\frac{1}{\gamma(1-\gamma^2)}\Big\{\Big[ (1-\gamma^2)\lambda_{H}+2\gamma\lambda_{R}             \Big]m_{R}^2-   \Big[ (1-\gamma^2)\lambda_{R}-2\gamma\lambda_{H}  \Big]m_{H}^2  \Big\},
       \label{vacuumreal}\\ 
     V_{R}=\frac{3}{2}\frac{\lambda_{R}(m^4_{H}-m^4_{R})-2\lambda_{H}m^2_{H}m^2_{R}}{\lambda^2_{R}+\lambda^2_{H}}; \label{vacuumreal1}\\                 
      V_{H}=\frac{3}{2}\frac{\lambda_{H}(m^4_{R}-m^4_{H})-2\lambda_{R}m^2_{H}m^2_{R}}{\lambda^2_{R}+\lambda^2_{H}}.
          \label{vacuumreal11}\\
          ({\rm det} {\cal H})_{ (V_{R})}=(\det {\cal H})_{V_H}=-\Big[ \frac{4\gamma(\gamma^2-1)}{3}\Big]^2(\lambda^2_{R}+\lambda^2_{H}) v^2_{0}w^2_{0} ;\label{vacuumreal2}
   \end{eqnarray}
 the strict negativity of $({\rm det} {\cal H})$ requires to fix to zero one of the vacuum expectation values; the choice $w_{0}=0$ leads to the following simplification
\begin{eqnarray}
 A)   \qquad\qquad\qquad  w_{0}=0, \quad \rightarrow \quad \frac{m^{2}_{H}}{m^{2}_{R}} =  \frac{(1-\gamma^{2})\lambda_{H}+ 2\gamma\lambda_{R}}{(1-\gamma^{2})\lambda_{R} -2\gamma\lambda_{H}}, \quad
v^{2}_{0} =  \frac{6am^{2}_{R}}{(1-\gamma^{2})\lambda_{R} -2\gamma\lambda_{H}}. \label{expvalue}
\end{eqnarray}
Similarly one has alternative the choice   
  \begin{eqnarray}
 B)   \qquad\qquad\qquad  v_{0}=0, \quad \rightarrow \quad \frac{m^{2}_{H}}{m^{2}_{R}}  =  \frac{(1-\gamma^{2})\lambda_{H}- 2\gamma\lambda_{R}}{(1-\gamma^{2})\lambda_{R} +2\gamma\lambda_{H}}, \quad
     w^{2}_{0}  =  \frac{6am^{2}_{R}}{(1-\gamma^{2})\lambda_{R} +2\gamma\lambda_{H}}. \label{expvalue2}
\end{eqnarray}                
Now we expand the potential around  the degenerate vacuum,
\begin{eqnarray}
     V(\psi+\psi_{0}, \overline{\psi}+\overline{\psi_{0}}) \!\! & = & \!\! \frac{am^{2}}{2} (\psi + \psi_{0})(\overline{\psi} + \overline{\psi_{0}}) + \frac{\lambda}{4!} (\psi + \psi_{0})^{2}(\overline{\psi}+\overline{\psi_{0}})^{2} \nonumber \\
     \!\! & = & \!\!- \frac{am^{2}}{2} \psi\overline{\psi} + \frac{\lambda}{4!} (\psi\overline{\psi})^{2} + \frac{\lambda}{4!} (\overline{\psi}^{2}_{0}\psi^{2} + \psi^{2}_{0}\overline{\psi}^{2}) + \frac{\lambda}{12} \psi\overline{\psi} (\psi_{0}\overline{\psi}+ \overline{\psi}_{0}\psi),
     \label{neve}
\end{eqnarray}
where
\begin{eqnarray}
     \overline{\psi}^{2}_{0} \psi^{2} + \psi^{2}_{0}\overline{\psi}^{2} \!\! & = & \!\! 2\cosh 2(\chi_{0}-\chi) \cos 2(\theta_{0}-\theta)  \Big\{ (\gamma^{4}+1)(w^{2}_{0}-v^{2}_{0})(w^{2}-v^{2}) -2\gamma^{2} [(v^{2}_{0}+3w^{2}_{0})w^{2} + (3v^{2}_{0}+w^{2}_{0})v^{2}]\Big\} \nonumber \\
     \!\! & & \!\! +8\gamma \sinh 2(\chi_{0}-\chi) \Big\{ (\gamma^{2}-1) \sin 2(\theta_{0}-\theta )(v^{2}_{0}v^{2}-w^{2}_{0}w^{2}) - (\gamma^{2}+1) \cos 2(\theta_{0}-\theta)v_{0}w_{0}(v^{2}+w^{2})\Big\} \nonumber \\
     \!\! & & \!\! +4 (\gamma^{4}-1) \cosh 2(\chi_{0}-\chi) \sin 2(\theta_{0}-\theta) v_{0}w_{0}(v^{2}-w^{2}) \nonumber \\
     \!\! & & \!\! +8 (\gamma^{2}+1) \cos 2(\theta_{0}-\theta) \Big[(\gamma^{2}+1) v_{0}w_{0} \cosh 2(\chi_{0}-\chi) + \gamma (v^{2}_{0}+w^{2}_{0}) \sinh 2(\chi_{0}-\chi)\Big]\underbrace {vw }\nonumber \\
     \!\! & & \!\! -4 (\gamma^{4}-1) (v^{2}_{0}-w^{2}_{0}) \sinh 2(\chi_{0}-\chi) \sin 2(\theta_{0}-\theta) \cdot \underbrace{vw }\nonumber \\
     \!\! & + & \!\! 2k \Big\{ 4\gamma (\gamma^{2}-1) \cosh 2(\chi_{0}-\chi) \cdot \cos 2(\theta_{0}-\theta) (w^{2}_{0}w^{2} - v^{2}_{0}v^{2}) \nonumber \\
     \!\! & & \!\! + \sinh 2(\chi_{0}-\chi) \sin 2(\theta_{0}-\theta) \Big[(\gamma^{4}+1)(w^{2}_{0}-v^{2}_{0}) (w^{2}-v^{2}) - 2\gamma^{2} (v^{2}_{0}+w^{2}_{0}) (v^{2}+w^{2}) \nonumber \\
     \!\! & & \!\!  -4 \gamma^{2} (v^{2}_{0}v^{2} +w^{2}_{0}w^{2})\Big] \nonumber \\
     \!\! & & \!\! +2 (\gamma^{4}-1) \sinh 2(\chi_{0}-\chi) \cos 2(\theta_{0}-\theta) v_{0}w_{0} (w^{2}-v^{2}) \nonumber \\
     \!\! & & \!\! -4 \gamma(\gamma^{2}+1) \cosh 2(\chi_{0}-\chi) \sin 2(\theta_{0}-\theta) v_{0}w_{0} (w^{2}+v^{2}) \Big\} \nonumber \\
     \!\! & + & \!\! 4k (\gamma^{2}+1) \sinh 2(\chi_{0}-\chi) \Big[2v_{0}w_{0} \sin 2(\theta_{0}-\theta) - (\gamma^{2}-1)(w^{2}_{0}-v^{2}_{0}) \cos 2(\theta_{0}-\theta)\Big]\underbrace{ vw }\nonumber \\
     \!\! & + & \!\! 8k \gamma (\gamma^{2}+1) \cosh 2(\chi_{0}-\chi) \sin 2(\theta_{0}-\theta) (v^{2}_{0}+w^{2}_{0})\underbrace{vw};
     \label{quacub1}
\end{eqnarray}  
plus higher order terms; the fields have the general form $\psi_{0} = [(\gamma +k)w_0+(i\gamma +j)v_0]e^{i\theta_0}e^{j\chi_0}$, and $\psi= [(\gamma +k)w+(i\gamma +j)v]e^{i\theta}e^{j\chi}$; note that the quadratic expression (\ref{quacub1}) has the Hermitian form with real and $k$-hybrid terms. 

In general the vanishing of the quadratic interaction terms  (both real and hybrid) in the Eq. (\ref{quacub1}) requires, for the v.e.v given in Eq. (\ref{expvalue}), that
\begin{eqnarray}
4v^2_{0}(\gamma^2+1)\Big[2\gamma \cos 2(\theta_0-\theta)-(\gamma^2-1)\sin 2(\theta_0-\theta)
\Big]\sinh2(\chi-\chi_0)=0; 
 \label{rotgamma}\\
 4kv^2_{0}(\gamma^2+1)\Big[2\gamma \sin 2(\theta_0-\theta)\cosh2(\chi-\chi_0)
+(\gamma^2-1)\cos 2(\theta_0-\theta)\sinh 2(\chi-\chi_0)\Big]=0;  
  \label{rotgamma1} \end{eqnarray}
it is evident that the above expressions are satisfied under the choice
\begin{equation}
\sin2(\theta-\theta_0)=0;\quad \sinh2(\chi-\chi_0)=0, \label{usualchoice}
\end{equation}
for $\gamma$ arbitrary; this gauge was used in \cite{1} for diagonalizing the mass matrix.  However there exists another possibility, which was unnoticed in \cite{1}, since the Eq. (\ref{rotgamma}) is satisfied
also under the restriction
\begin{equation}
\tan2(\theta_0-\theta)=\frac{2\gamma}{\gamma^2-1};
\label{diagonal2}
\end{equation}
and then the Eq. (\ref{rotgamma1}) implies that 
\begin{equation}
\tanh 2(\chi_0-\chi)=-\Big(\frac{2\gamma}{\gamma^2-1}\Big)^2;
\label{diagonal3}
\end{equation}
therefore, the remaining quadratic terms in the Eq. (\ref{quacub1}) are determined by the expressions
\begin{eqnarray}
\cosh2(\chi_0-\chi)\cos 2(\theta_0-\theta)=s\frac{(\gamma^2-1)^3}{(\gamma^2+1)^2\sqrt{P^+P^-}},\quad s=\pm 1; \nonumber \\
\sinh 2(\chi_0-\chi)\sin 2(\theta_0-\theta) = l\frac{8\gamma^3}{(\gamma^2+1)^2\sqrt{P^+P^-}}; \quad l=\pm 1;
\label{diagonal4}
\end{eqnarray}
where $sl=-1$, and the polynomials $P^+$, and $P^-$ are defined below in Eqs. (\ref{positivequan}).

Similarly for the v.e.v. corresponding to case B) and given in Eq. (\ref{expvalue2}), the expressions analogous to those in Eqs. (\ref{diagonal2}), and (\ref{diagonal3}) are essentially the same with a change of sign in the first one,
\begin{eqnarray}
\tan2(\theta_0-\theta)=-\frac{2\gamma}{\gamma^2-1};
\quad
\tanh 2(\chi_0-\chi)=-\Big(\frac{2\gamma}{\gamma^2-1}\Big)^2;
\label{diagonal5}
\end{eqnarray}
the expressions (\ref{diagonal4}) remain valid but with the restriction $sl=1$.

\section{The case $(v_0\neq 0,w_0=0)$, and $\lambda_{R}=\lambda_{H}$:}
\label{gammano1}

Due to the proliferation of parameters in the hypercomplex model, namely $(m^2_R, m^2_H, \lambda_R, \lambda_H;\gamma)$, the restriction $\lambda_R=\lambda_H$ was considered in \cite{1} in order to simplify the analysis; if one enforces the constraint $m^2_H=m^2_R$, then necessarily $\gamma=0$, recovering the usual $U(1)$ gauge theory.
Hence, from the expressions  (\ref{expvalue}) we find that the mass ratio is the physically dimensionless quantity defined in terms of the $\gamma$-parameter; similarly the vacuum expectation value for the field $v$ is defined as a $\gamma$-deformation of the conventional value $\frac{-6m^2_R}{a\lambda_R}$,
\begin{eqnarray}
     \frac{m^{2}_{H}}{m^{2}_{R}} = \frac{P^-(\gamma)}{P^+(\gamma)}, \qquad v^{2}_{0} = \frac{6m^{2}_{R}}{-a\lambda_{R}P^+(\gamma)}, 
\quad P^+(\gamma)\equiv \gamma^{2}+2\gamma -1; \quad P^-(\gamma)\equiv P^+(-\gamma);
     \label{positivequan}
\end{eqnarray}
and hence the expressions (\ref{falsev}), and (\ref{falsevv}) for the Hessians at the zero-energy point will take the form
\begin{eqnarray}
\det {\cal H}_{ (V_{R})}=m^4_R\Big[\frac{(\gamma^2+1)}{P^+(\gamma)}\Big]^2\widehat{P}_+(\gamma), \quad 
\det {\cal H}_{ (V_{H})}=m^4_R\Big[\frac{(\gamma^2+1)}{P^+(\gamma)}\Big]^2\widehat{P}_-(\gamma); \nonumber\\
\widehat{P}_+(\gamma)\equiv \gamma^4+4\gamma^3-6\gamma^2-4\gamma+1, \quad  \widehat{P}_-(\gamma)\equiv\widehat{P}_+(-\gamma);
\label{P}
\end{eqnarray}
the values of $\gamma$ must be restricted to the interval $(\gamma_H, -\gamma_H)$, in which the above expressions
take positive values, and a stable vacuum is induced; 
the limits of the interval correspond to roots of $\det {\cal H}_{ (V_{H})}$, and $\det {\cal H}_{ (V_{R})}$ in the above expressions \cite{1}
\begin{eqnarray}
\gamma\in (\gamma_H, -\gamma_H), \quad \gamma_H=1+\sqrt{2}-\sqrt{2(2+\sqrt{2})}\approx -0.1989; \quad
\quad
\widehat{P}_-(\gamma_H)=0=\widehat{P}_+(-\gamma_H).
\label{roots}
\end{eqnarray}
Hence the potentials take the form
\begin{eqnarray}
V_{R}(v,w;\gamma)= \frac{a}{2}m^2_{R}\Big[P^{v}_{R}v^2+P^{w}_{R}w^2\Big]+ \frac{\lambda_R}{4!} \Big[ \widehat{P}_+(\gamma)v^4+\widehat{P}_-(\gamma)w^4+2(\gamma^2+1)^2v^2w^2
\Big]; 
     \label{VRS} \\
V_{H}(v,w;\gamma)= \frac{a}{2}m^2_R\Big[P^{v}_{H}v^2+P^{w}_{H}w^2\big]+
\frac{\lambda_R}{4!} \Big[ \widehat{P}_-(\gamma)v^4+\widehat{P}_+(\gamma)w^4+2(\gamma^2+1)^2v^2w^2
\Big]; 
\label{VHS}    
\end{eqnarray} 
 where the polynomials are defined as
 \begin{eqnarray}
 P^{v}_{R}=\frac{\widehat{P}_+(\gamma)}{P^+(\gamma)}, \quad P^{v}_{H}=\frac{\widehat{P}_-(\gamma)}{P^+(\gamma)}, \quad P^{w}_{R}= \frac{(\gamma^2+1)^2}{P^+(\gamma)}= P^{w}_{H};
\label{polymass}
\end{eqnarray}
These potentials were analyzed in \cite{1}, and have basically the profile shown in the figure \ref{fig11}, as functions on $(v,w)$, and for 
some value of $\gamma$ in the range $(\gamma_H, -\gamma_H)$. 

As a complement of the results obtained in \cite{1}, we analyze here the effective self-interactions, and the effective mass terms, in order to establish a duality between the potentials.
A field with strong self-interactions is not manipulable with a perturbative expansion in the self-coupling constant; on the other hand, if such a constant is too small, then the contribution of quantum corrections to the effective potential would be dominant. Since the mass terms and self-coupling constants appear in the present scheme as one-parameter flows in $\gamma$, the duality will allow us to connect  weak self-interaction with strong-self-interaction regimes.

\begin{figure}[H]
  \begin{center}
  \includegraphics[width=.5\textwidth]{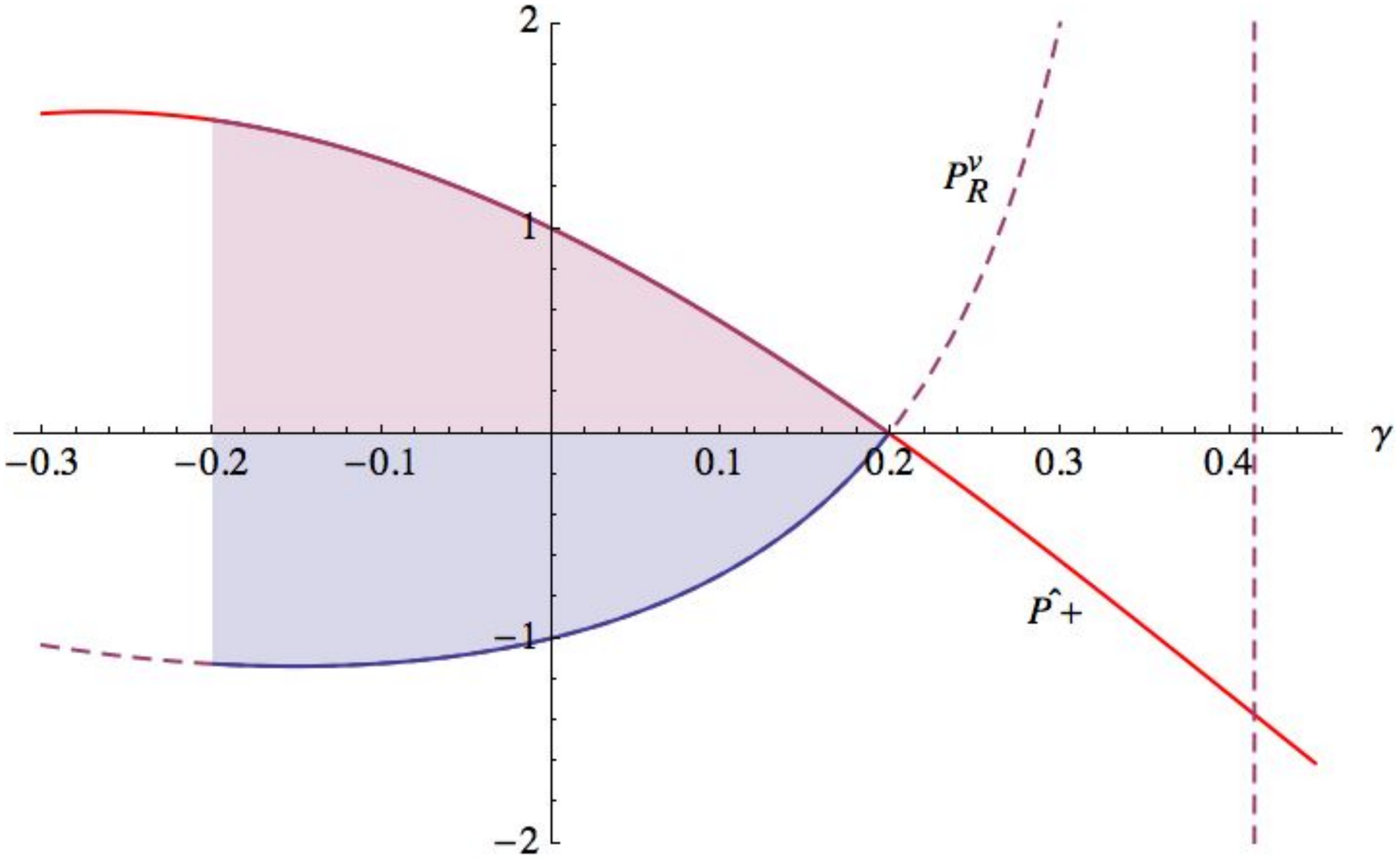}
\end{center}
\caption{The effective mass and the self-interaction of the field $v$ in the potential $V_R$, in the range $(\gamma_H, -\gamma_H)$.}
\label{int1}
\end{figure}

In relation to the potential $V_R$, the figure \ref{int1} shows that the right limit $\gamma\rightarrow -\gamma_{H}$ is the limit of masslessness and weak self-interaction for the field $v$; similarly the left limit $\gamma\rightarrow \gamma_{H}$, corresponds to a limit for a massive and strong self-interacting  field $v$. 

\begin{figure}[H]
  \begin{center}
  \includegraphics[width=.5\textwidth]{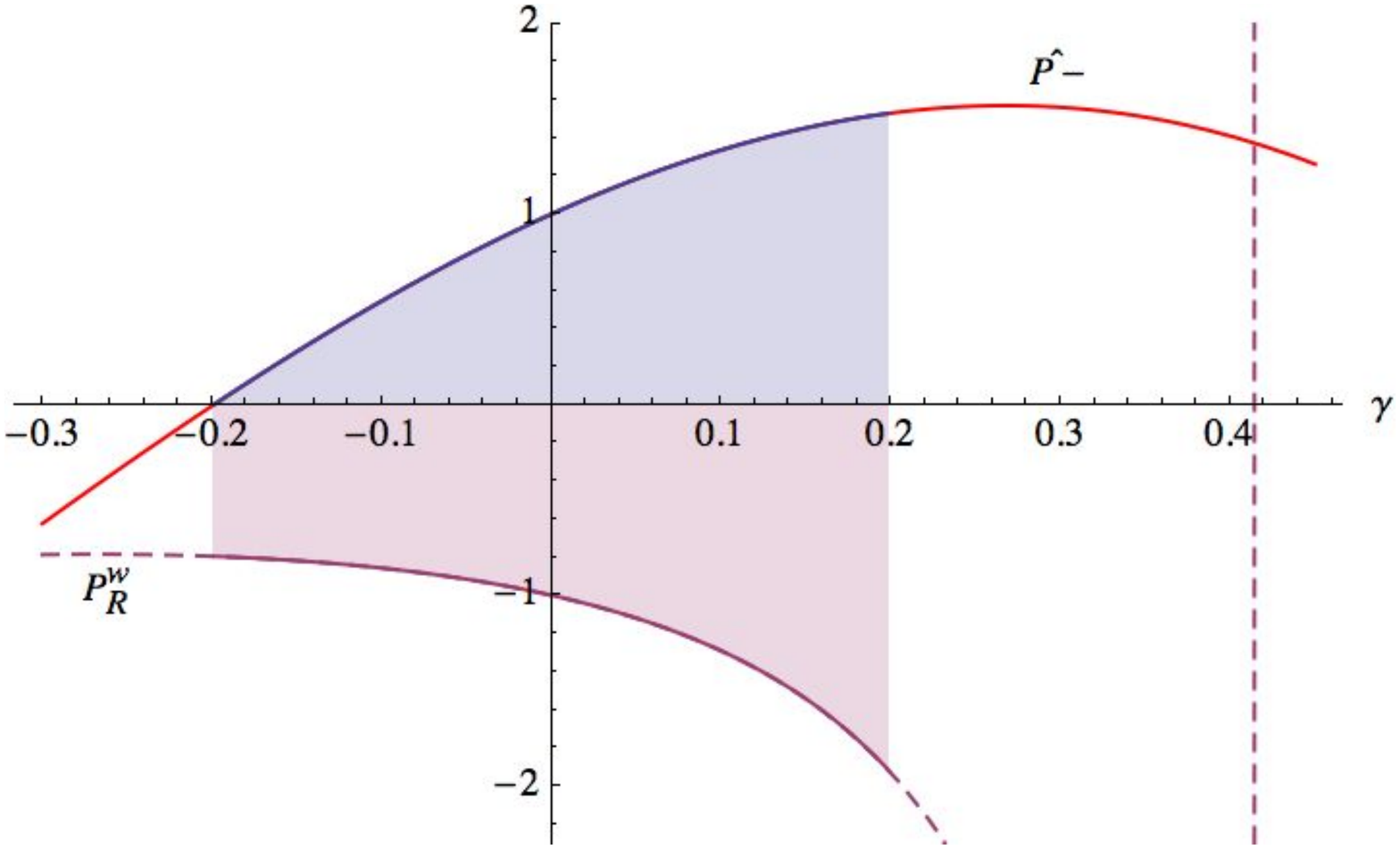}
\end{center}
\caption{The effective mass and the self-interaction of the field $w$ in the potential $V_R$, and in the range $(\gamma_H, -\gamma_H)$.}
\label{int2}
\end{figure}

\begin{figure}[H]
  \begin{center}
  \includegraphics[width=.5\textwidth]{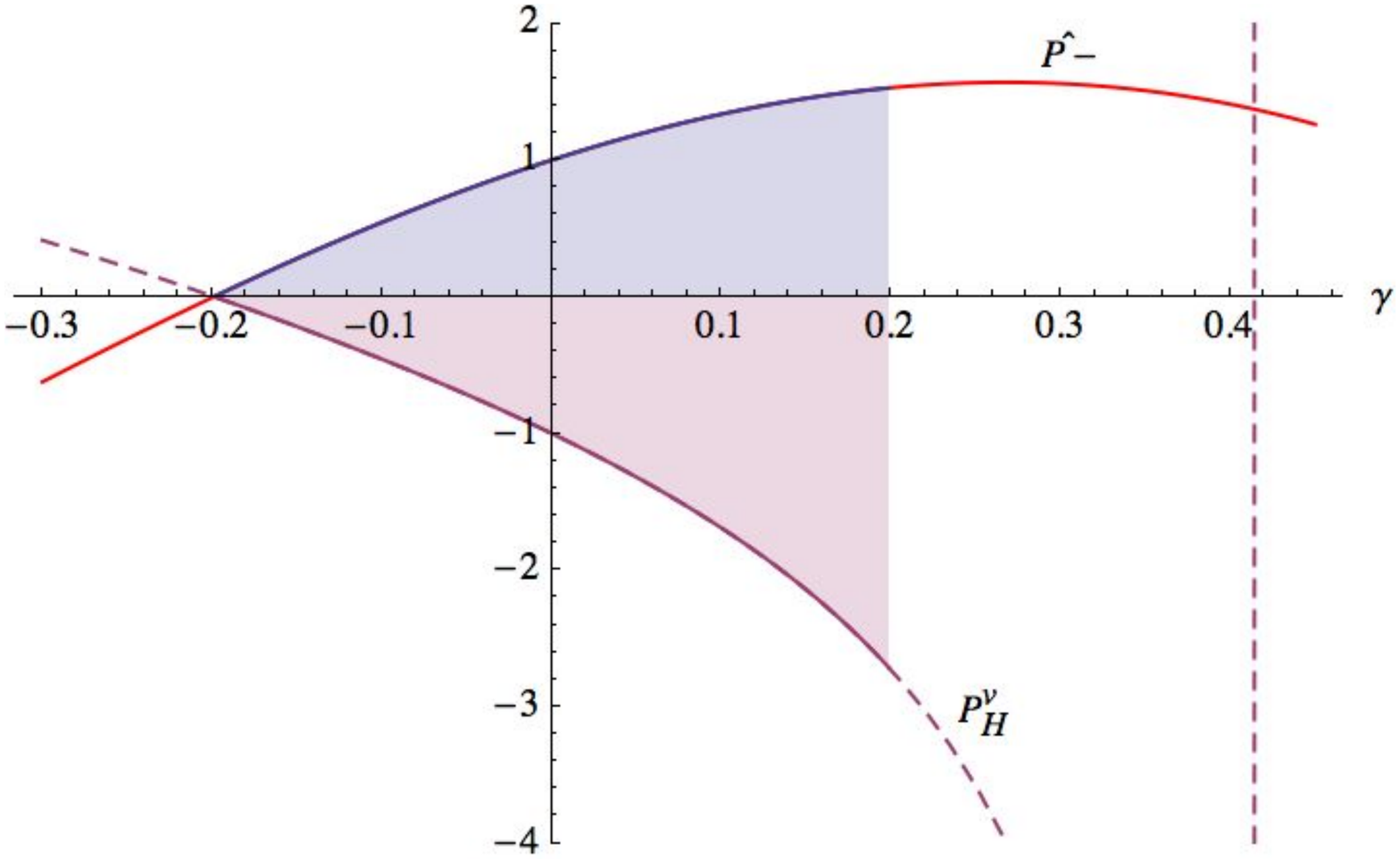}
\end{center}
\caption{The effective mass and the self-interaction of the field $v$ in the potential $V_H$, and in the range $(\gamma_H, -\gamma_H)$.}
\label{int3}
\end{figure}

Furthermore, in relation to the same potential $V_R$, the figure \ref{int2} shows that the right limit $\gamma\rightarrow -\gamma_{H}$ corresponds now to the massive and strong self-interaction
limit for the field $w$; the limit $\gamma\rightarrow \gamma_{H}$ is for a weak self-interacting, and a relatively light field. Hence, at the {\bf right} limit 
$-\gamma_H$ we have a mix of the strong-self-interaction regime for $v$, with a weak regime for the field $w$, and conversely at the {\bf left} limit $\gamma_H$. 
Now, in relation to the potential $V_H$, the figure \ref{int3} shows that the regimes for the field $v$ are interchanged at the limits $\pm \gamma_H$; therefore, 
a direct comparison between the figure \ref{int1}, and figure \ref{int3}, shows that, with respect to the effective mass and the self-interaction of the field $v$, the potentials $V_R$, and  $V_H$ are dual to each other at those limits.  Since the self-interaction terms for each field in the potential $V_H$ are described by the respective mirror polynomials ($\hat{P}_+, \hat{P}_-$), then 
the roles of the fields are interchanged, establishing thus the duality between the potentials. 
The cross-coupling term $v^2w^2$ has the same effective expression for both potentials, and we have then self-dual terms under the duality; similarly the (input)  mass term for $w$ is the same in both potentials. In general the mass polynomials in the figures (\ref{int1}), (\ref{int2}), and (\ref{int3}) will be modified after the spontaneous symmetry breaking; the quartic self-interaction terms will remain intact (see Eq. \ref{neve}), and thus the duality between the weak and strong regimes of the fields will remain unchanged.

As already mentioned in the introduction, the spontaneous symmetry breaking was studied in \cite{1}  by using the gauge (\ref{usualchoice}); now we explore the gauge 
(\ref{diagonal2}), and (\ref{diagonal3}),  by replacing the expressions (\ref{diagonal4}), and (\ref{diagonal5}) into the quadratic form 
(\ref{quacub1}), that yields a canonical expression for the mass terms,
\begin{equation}
      -\frac{am^2}{2}\psi\overline{\psi} + \frac{\lambda}{4!}(\overline{\psi}^{2}_{0} \psi^{2} + \psi^{2}_{0} \overline{\psi}^{2} ) 
= am^2_{R}\Big[S_R^v(\gamma) v^2+S_R^w(\gamma) w^2+kS_H^v(\gamma)v^2+kS_H^w(\gamma)w^2\Big];
\label{pprime} 
\end{equation} 
where
\begin{eqnarray}
S_R^v(\gamma;l,s)\equiv -\frac{1}{2}P^v_R -\frac{1}{2}\frac{s(\gamma^2-1)^3P^v_R-8l\gamma^3P^v_H}{(\gamma^2+1)^2\sqrt{P^+P^-}},\label{s1}\\
S^w_R(\gamma;l,s)\equiv -\frac{1}{2}P^w_R+\frac{1}{2}\frac{s(\gamma^2-1)^3-8l\gamma^3}{P^+\sqrt{P^+P^-}},\label{s2}\\
S_H^v(\gamma;l,s)\equiv -\frac{1}{2}P^v_H -\frac{1}{2}\frac{s(\gamma^2-1)^3P^v_H+8l\gamma^3P^v_R}{(\gamma^2+1)^2\sqrt{P^+P^-}},\label{s3}\\
S^w_H(\gamma;l,s)\equiv -\frac{1}{2}P^w_R+\frac{1}{2}\frac{s(\gamma^2-1)^3+8l\gamma^3}{P^+\sqrt{P^+P^-}},\label{s4}
\end{eqnarray}
and we have used the fact that  $(v_0\neq 0,w_0=0)$. These expressions depend fully on $\gamma$, remaining to fix the pair $(l,s)$; with respect to generating a mass hierarchy, the two possibilities will show different physical properties.

\subsection{$s=1, l=-1$}
\begin{figure}[H]
  \begin{center}
  \includegraphics[width=.5\textwidth]{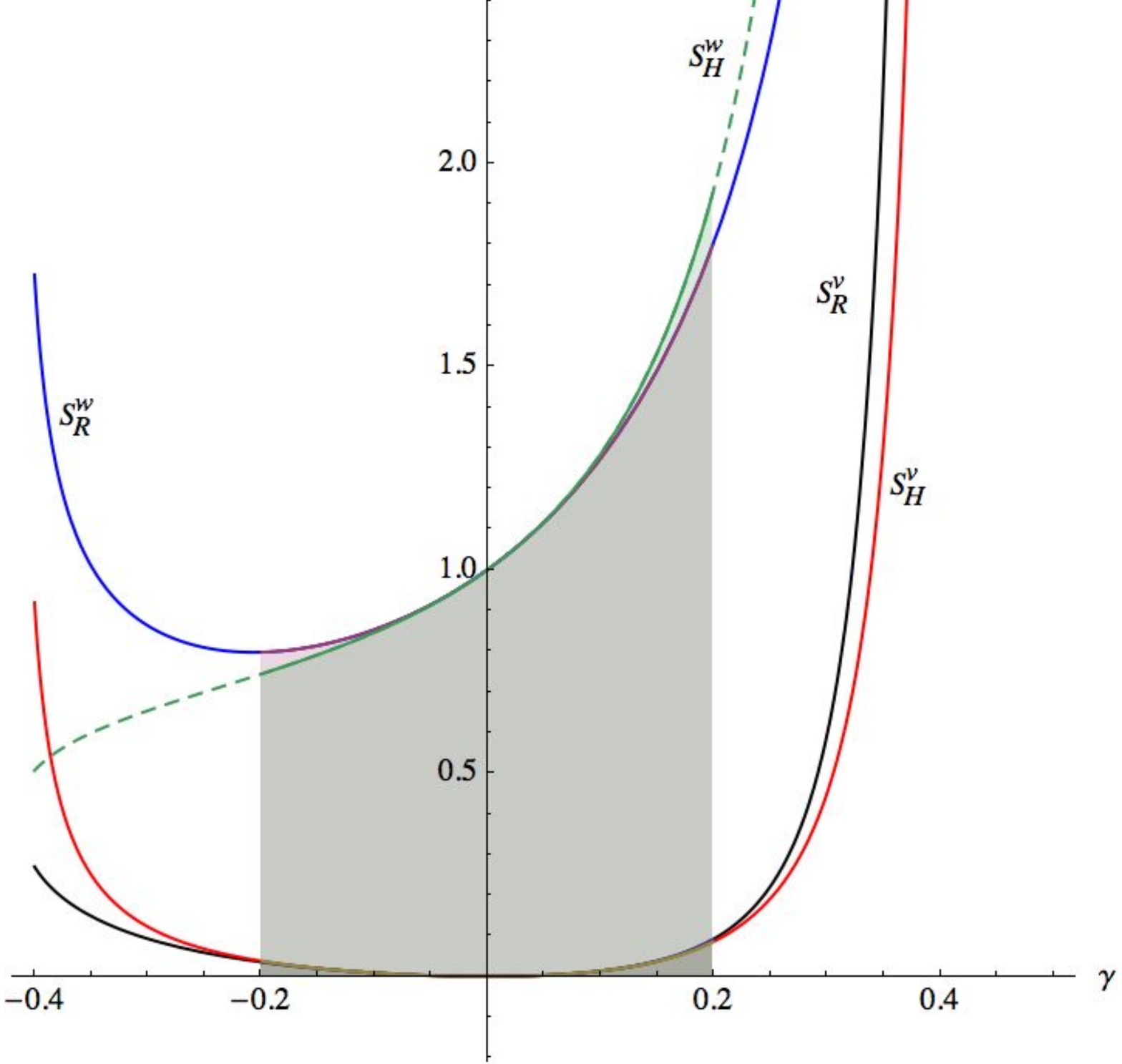}
\end{center}
\caption{The peculiar feature of this spectrum, the heaviness of $w$, and the lightness of $v$; in the shadowed region all polynomials are positive.}
\label{sr}
\end{figure}

For this choice, the polynomials (\ref{s1})-(\ref{s4}) are shown in the figure \ref{sr}. Hence, something remarkable happens since the mass polynomials for the field $v$ are separated from the mass polynomials
for the field $w$, inducing a hierarchy; note that the input mass polynomials $(P^v_R, P^v_H, P^w_R, P^w_R)$ illustrated in the figures
\ref{int1}, \ref{int2}, and \ref{int3}, show no hierarchy. Once such a separation is made, there is a fine splitting between the mass polynomials for each field; the polynomials for each field coincide at $\gamma=0$, and the splitting is maximum at the limits $\pm\gamma_H$.

Quantitatively we have that, at the limits of the interval:

{\bf the left limit $\gamma_H$:}
\begin{eqnarray}
\gamma_H: \quad S^v_R(s=1,l=-1; \gamma_{H}) 	\approx 0.0350, \quad S^v_H(s=1,l=-1; \gamma_{H}) \approx 0.0375; \nonumber\\
S^w_R(s=1,l=-1; \gamma_{H}) 	\approx 0.7974, \quad S^w_H(s=1,l=-1; \gamma_{H}) 	\approx 0.7444; 
\label{s5}
\end{eqnarray}
therefore, in $[m^2_R]$ units, the masses for each field are of the same order, and the hierarchy is of order $10^2$ in favor of $w$. 
These mass polynomials generated by spontaneous symmetry breaking must be complemented with the interaction regimes described 
in the figures \ref{int1}, \ref{int2}, and \ref{int3}.

Similarly, we have for the other limit,

{\bf the right limit $-\gamma_H$:}
\begin{eqnarray}
-\gamma_H: \quad S^v_R(s=1,l=-1; -\gamma_{H}) \approx 0.0905, \quad S^v_H(s=1,l=-1; -\gamma_{H}) \approx 0.0845; \nonumber\\
 S^w_R(s=1,l=-1; -\gamma_{H}) \approx 1.7971, \quad S^w_H(s=1,l=-1; -\gamma_{H}) \approx 1.9251.
\label{s6}
\end{eqnarray}
The exact values for these expressions are available, but correspond to very complicated irrational expressions; we have displayed the approximated values up to certain decimal places for practical purposes, and they should be sufficiently accurate with the idea of defining a hierarchy.  The exact value of $\gamma_{H}$ is given in the expression (\ref{roots}), and we can give an exact value  as example; the simplest exact expression between them is $S^v_R$  in Eq. (\ref{s6}),
\begin{equation}
S^v_R(s=1,l=-1; -\gamma_{H}) = \frac{31 + 23 \sqrt{2} - 16 \sqrt{2 + \sqrt{2}} - 
 13 \sqrt{2 (2 + \sqrt{2})}}
 {\Big(2 + \sqrt{2} - 2 \sqrt{2 + \sqrt{2}}\Big) \sqrt{
 15 + 11 \sqrt{2} - 8 \sqrt{2 + \sqrt{2}} - 6 \sqrt{2 (2 + \sqrt{2})}}}.
 \end{equation}
Furthermore, the polynomial $S^w_R$ has a global minimum at $\gamma\approx -0.2065$, however lies out of the interval $(\gamma_H,-\gamma_H)$, and it is not a criterion for fixing the value of $\gamma$.

We can arbitrarily fix $\gamma$ as close to zero as wanted, and thus the field $v$ tends to be light, and the field $w$ will have masses close to 1 in $[m^2_R]$ units.
The values of $\gamma$ at the limits $\pm\gamma_H$ are of order $10^{-1}$, and for the purposes of comparison, we can take $\gamma$ of order $10^{-2}$, and determine the order of magnitude of the mass hierarchy generated; hence, we have that 
\begin{eqnarray}
S^v_R(\gamma\approx 10^{-2}) \approx 10^{-4}, \qquad S^v_H(\gamma\approx 10^{-2}) \approx 10^{-4}; \nonumber\\
S^w_R(\gamma\approx 10^{-2}) \approx 1.02061..., \qquad S^w_H(\gamma\approx 10^{-2}) \approx 1.02062...; 
\label{largeh}
\end{eqnarray}
with a hierarchy of order $10^{4}$, which must be compared with the order $10^2$ in the expressions (\ref{s5}). Roughly speaking, if $\gamma\approx 10^{-n}$, then the hierarchy will be of order $10^{2n}$ in favor of the field $w$. Thus, by assigning Planck scale masses to the field $w$ and electroweak masses for the field $v$ we generate the desired hierarchy by choosing $n=8$ in this scheme.

However, this fine-tuning for the $\gamma$-parameter is very artificial, and one would like to achieve it naturally;
we consider now the another possibility for the pair $(l,s)$, which will show, surprisingly, different qualities.

\subsection{$s=-1, l=1$}

\begin{figure}[H]
  \begin{center}
  \includegraphics[width=.5\textwidth]{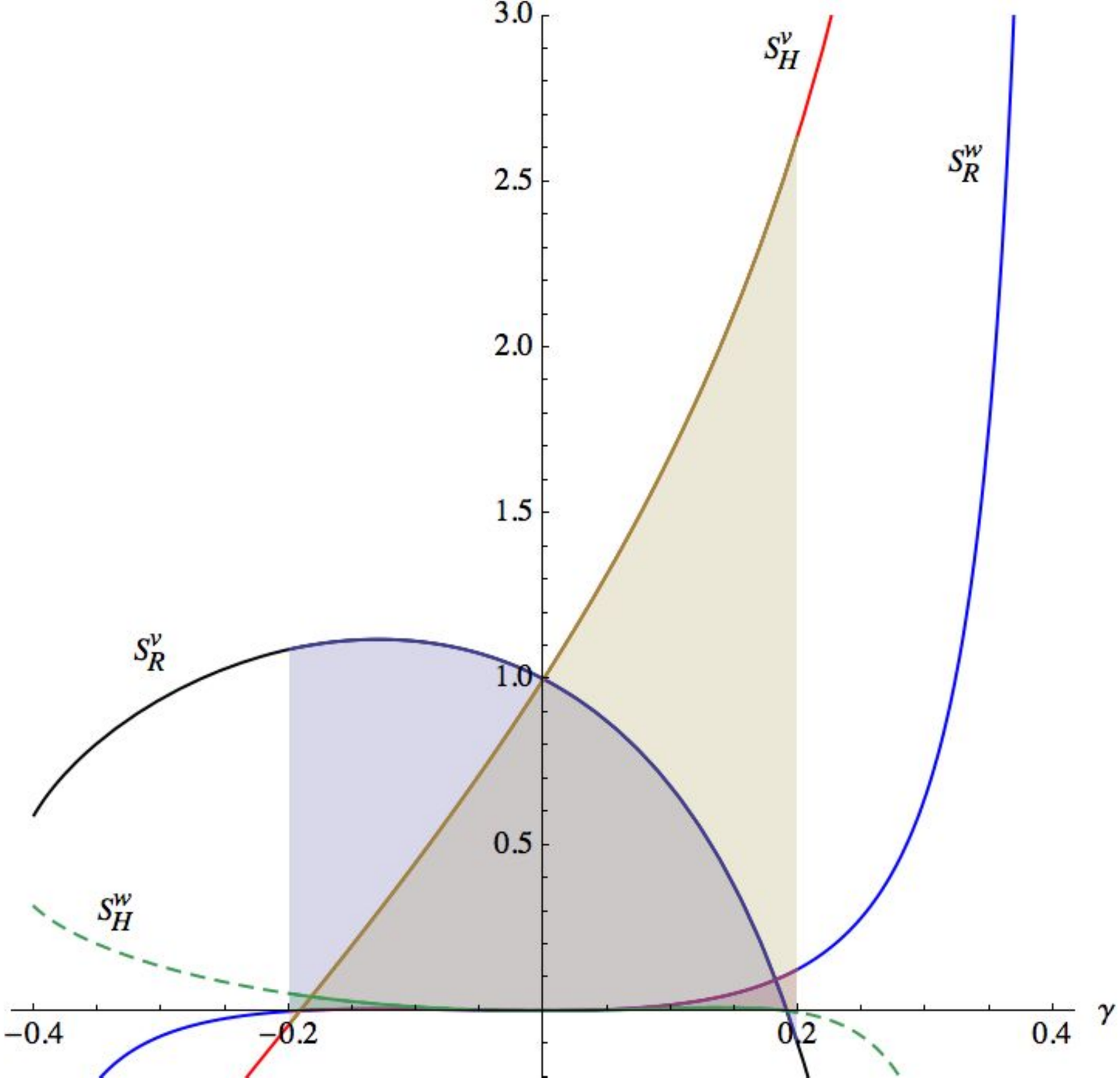}
\end{center}
\caption{The mass polynomials in Eq. (\ref{pprime}), for $s=-1$, and $l=1$; in the shadowed region all polynomials are positive.}
\label{ssr}
\end{figure}

For this case the polynomials are shown in the figure \ref{ssr}; the fine splitting only is present for the polynomials of the field $w$; however, the polynomials will show 
critical points that represent possible criteria for fixing the value of $\gamma$.
First, at the limits of the interval we have that,

{\bf The left limit $\gamma_H$:}
\begin{eqnarray}
 \gamma_{H}: \quad S^v_R(s=-1,l=1; \gamma_{H}) 	\approx 1.0902, \quad S^v_H(s=-1,l=1; \gamma_{H}) \approx -0.0375;\nonumber \\
 S^w_R(s=-1,l=1; \gamma_{H}) 	\approx  -0.001763, \quad S^w_H(s=-1,l=1; \gamma_{H}) 	\approx 0.05126; 
 \label{s7}
 \end{eqnarray}
 {\bf The right limit $-\gamma_H$:}
  \begin{eqnarray}
- \gamma_{H}: \quad    S^v_R(s=-1,l=1; -\gamma_{H}) \approx -0.0905, \quad S^v_H(s=-1,l=1; -\gamma_{H}) \approx 2.6320;\nonumber\\
S^w_R(s=-1,l=1; -\gamma_{H}) \approx 0.1237, \quad S^w_H(s=-1,l=1; -\gamma_{H}) \approx -0.00425;\label{s8}
\end{eqnarray}
and we see that some polynomials take negative values, and qualitatively we do not consider such limits as physically viable for fixing the value of $\gamma$.

\begin{figure}[H]
  \begin{center}
  \includegraphics[width=.5\textwidth]{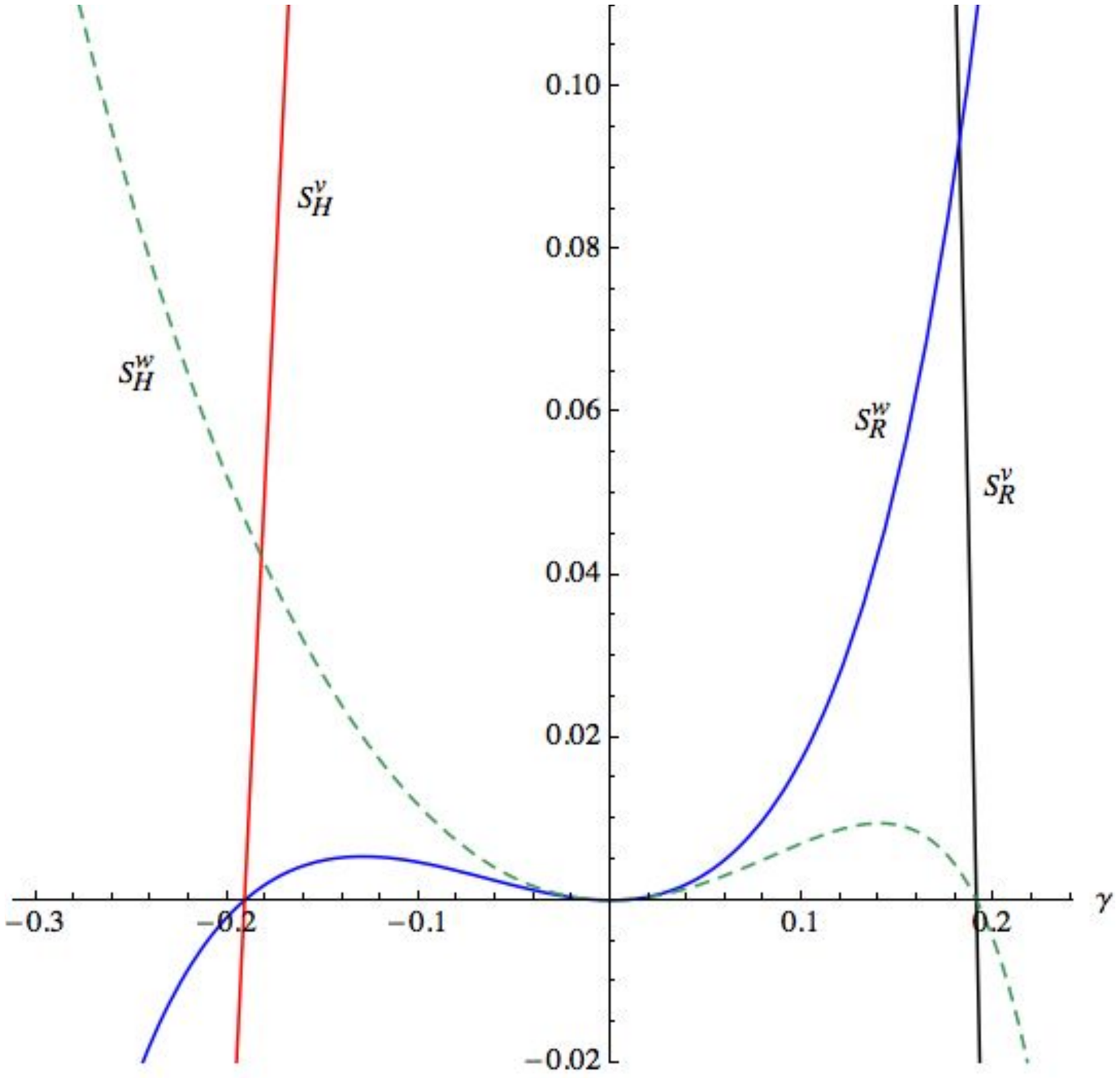}
\end{center}
\caption{ A zoom of the figure \ref{ssr}.}
\label{zoomssr}
\end{figure}

Now, $S^v_R$, and $ S^w_H$, have a common root at $\gamma\approx 0.19106<-\gamma_H$; although it is not visible in the figure \ref{ssr}, the zoom in the figure \ref{zoomssr} shows it.

{\bf A common root for $S^v_R$, and $ S^w_H$:}
\begin{eqnarray}
S^v_R(s=-1,l=1; \gamma\approx 0.19106) =0=S^w_H(s=-1,l=1; \gamma\approx 0.19106),\nonumber \\
 S^v_H(s=-1,l=1; \gamma\approx 0.19106) \approx 2.5379, \quad S^w_R(s=-1,l=1; \gamma\approx 0.19106) \approx 0.1085;\label{s9}
\end{eqnarray}
therefore, at this point we have a  $k$-massive field $v$, and a real massive field $w$; the hierarchy is of order $10$, in favor of $v$. At this point,
we have additionally that $\hat{P}_{+}(\gamma\approx 0.19106)\approx 4.596\times 10^{-2}$, and $\hat{P}_{-}(\gamma\approx -0.12966)\approx 1.5186$; hence, in relation to the potential $V_{R}$, we can consider that the field $v$ is weakly self-interacting with respect to the field $w$, and conversely in the potential $V_H$.

Similarly  $S^v_H$, and $S^w_R$, have a common root at the opposite value considered above.

{\bf A common root for $S^v_H$, and $S^w_R$:}
\begin{eqnarray}
 S^v_H(s=-1,l=1; \gamma\approx -0.19106) =0=S^w_R(s=-1,l=1; \gamma\approx -0.19106);\nonumber\\
 S^v_R(s=-1,l=1; \gamma\approx -0.19106) \approx 1.0965, \quad  S^w_H(s=-1,l=1; \gamma\approx -0.19106) \approx 0.04688;
 \label{s10}
\end{eqnarray}
however the hierarchy is of order $10^2$. In relation to the self-interaction regimes, the roles of the fields $v$, and $w$ are interchanged with respect to the previous case, without changing the order of magnitude of the self-interactions.

$S^v_R$ is increasing monotonically from its (negative) value at $-\gamma_{H}$, passing through its root, and then to a maximum value at 
$\gamma\approx -0.12966>\gamma_H$;

{\bf Global maximum for $S^v_R$:}
\begin{eqnarray}
S^v_R (s=-1,l=1; \gamma\approx -0.12966)\approx 1.11963; \quad
S^v_H (s=-1,l=1; \gamma\approx -0.12966)\approx 0.29857;\nonumber\\
S^w_R (s=-1,l=1; \gamma\approx -0.12966)\approx  0.00538; \quad
S^w_H (s=-1,l=1; \gamma\approx -0.12966)\approx 0.02018; 
\label{smax}
\end{eqnarray}
and then decreases to its value at $\gamma_{H}$. At this point, there is a hierarchy of order 10 in relation to the masses of the same field $v$, and a hierarchy of orders $10^2$, and  $10^3$ in relation to the masses of the field $w$. In this case we have that  $\hat{P}_{+}(\gamma\approx -0.12966)\approx 1.4093$, and $\hat{P}_{-}(\gamma\approx 0.19106)\approx 3.894\times 10^{-1}$.

In the figure \ref{zoomssr} we see that the polynomial $S^w_H$ has at minimum at $\gamma=0$,  which is also a root; additionally it has a local maximum at $\gamma \approx 0.14042$,

{\bf Local maximum for $S^w_H$:}
\begin{eqnarray}
S^w_H(s=-1,l=1; \gamma \approx 0.14042 )\approx 0.00947; \quad
  S^w_R(s=-1,l=1; \gamma \approx 0.14042) \approx 0.04319; \nonumber\\
  S^v_R (s=-1,l=1; \gamma\approx 0.14042)\approx 0.44226, \quad S^v_H (s=-1,l=1; \gamma\approx 0.14042)\approx 2.01746;
\label{ssmax}
    \end{eqnarray}
with an interchange of the roles of real mass and hybrid $k$-mass for the fields, the structure of the hierarchy is basically the same described in the expressions (\ref{smax}); the effective self-interaction polynomials have in effect close values $\hat{P}_{+}(\gamma\approx 0.14042)\approx  3.314\times 10^{-1}$, and $\hat{P}_{-}(\gamma\approx 0.14042)\approx1.4327$, but with an interchange of roles.
     
\begin{figure}[H]
  \begin{center}
  \includegraphics[width=.5\textwidth]{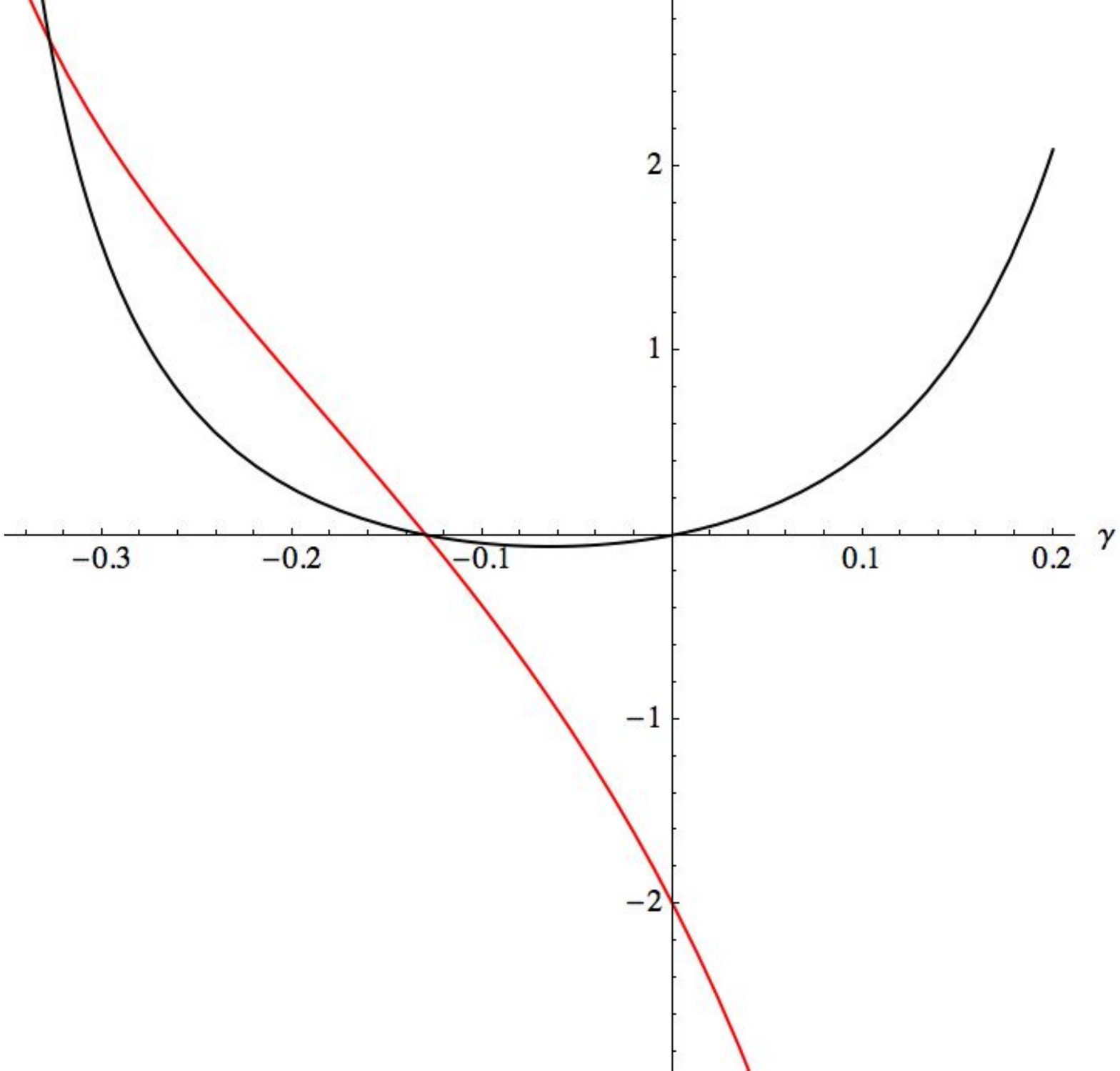}
\end{center}
\caption{The derivatives $\frac{d}{d\gamma}S^v_R$ (red), and $\frac{d}{d\gamma}S^w_R$ (black), which vanish at $\gamma\approx -0.12966$, the maxima described in Eq. (\ref{smax}), and Eq. (\ref{sssmax}).}
\label{derivative}
\end{figure}

Similarly the polynomial $S^w_R$ has at minimum at $\gamma=0$, and a local maximum at $\gamma \approx -0.12933$,

{\bf Local maximum for $S^w_R$:}
\begin{eqnarray}
S^w_R(s=-1,l=1; \gamma \approx -0.12933 )\approx 0.00538; \quad
  S^w_H(s=-1,l=1; \gamma \approx -0.12933 ) \approx 0.02007; \nonumber\\
  S^v_R (s=-1,l=1; \gamma\approx -0.12933 )\approx  1.11963, \quad S^v_H (s=-1,l=1; \gamma\approx -0.12933 )\approx 0.30022;
\label{sssmax}
    \end{eqnarray}
this case is qualitatively equivalent to the case described in the expressions (\ref{smax}), due to the fact that the maximum for $S^v_R$, and the local maximum
for $S^w_R$ coincide; the values of $\gamma$ and the values of the polynomials are distinguishable
to certain decimal place, due to the rough of our numerical analysis, and that we are approaching to the same point with different functions, and with different derivatives;  however, the derivatives are shown in the figure \ref{derivative}, which shows that they vanish at the same value of $\gamma$. Such a coincidence may reinforce
the choice of this critical point for establishing a hierarchy. Note that there exists another crossing point for the derivatives, but out of the range $(\gamma_{H},-\gamma_H)$. 
Another possible criterion is the choice of the crossing point of the polynomials $S^v_H$, and $S^w_H$, at $\gamma\approx -0.18224>\gamma_H$; therefore the fields $v$, and $w$ will have the same $k$-mass;

{\bf Crossing point of $S^v_H$, and $S^w_H$:}
\begin{eqnarray}
S^v_H( \gamma\approx -0.18224)\approx 0.04224 \approx S^w_H( \gamma\approx -0.18224);\nonumber\\
S^v_R( \gamma\approx -0.18224)\approx 1.10259,\quad  S^w_R( \gamma\approx -0.18224)\approx 0.00161;
\label{1crossing}
\end{eqnarray}
hence, the hierarchy is of order $10^3$  between the real masses of the fields, and of order $10^2$ in relation to the $k$-masses. In this case 
we have that  $\hat{P}_{+}(\gamma\approx -0.18224)\approx  1.5065$, and $\hat{P}_{-}(\gamma\approx -0.18224)\approx 1\times10^{-1}$.

Similarly, the crossing point between the polynomials $S^v_R$, and $S^w_R$, both associated with the real masses, is at the inverse value considered above,  $\gamma\approx 0.18224<-\gamma_H$,

{\bf Crossing point of $S^v_R$, and $S^w_R$:}
\begin{eqnarray}
S^v_R( \gamma\approx 0.18224)\approx 0.09336 \approx S^w_R( \gamma\approx 0.18224);\nonumber\\
S^v_H( \gamma\approx 0.18224)\approx 2.4371,\quad  S^w_H( \gamma\approx 0.18224)\approx 0.00357;
\label{2crossing}
\end{eqnarray}
in this case, the hierarchy is basically the same described in the expressions (\ref{1crossing}), but interchanging the roles between the real and $k$-masses. Note that in the last two cases the hierarchy separates the kind of masses.
\section{The case $(v_0\neq 0,w_0=0)$, and $m_{R}=m_{H}$}
\label{no1}
\begin{figure}[H]
  \begin{center}
  \includegraphics[width=.5\textwidth]{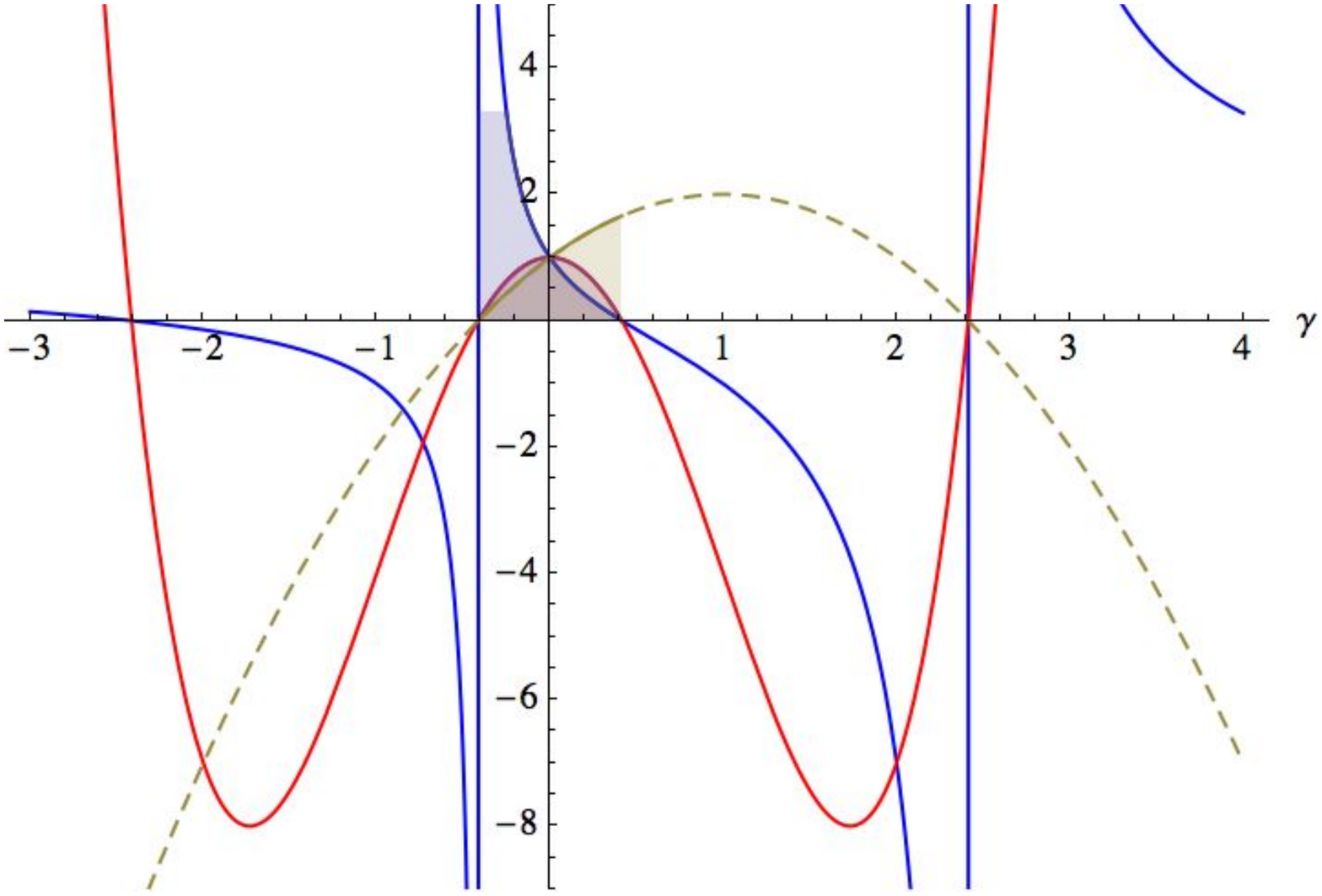}
\end{center}
\caption{The continuous red curve represents essentially $\det {\cal H}(0,0)$; 
the continuous blue curve represents the polynomial $\frac{P^+}{P^-}$, the ratio in  Eq. (\ref{positivequan}); the dashed curve represents the polynomial $-P^-$, Eq. (\ref{ineq4}); the shadowed region shows the interval where all polynomials are positive.}
\label{poly4}
\end{figure}

With respect to the parameters that define the model, and as opposed to the restrictions considered in section \ref{gammano1}, we consider now the case with the mass restriction 
 $m_{H} = m_{R} $; hence the expressions (\ref{expvalue}) reduce to
\begin{equation}
     \frac{\lambda_{H}}{\lambda_{R}} =\frac{P^+}{P^-},    
       \quad v^{2}_{0} = \frac{6m^{2}_{R}}{(\gamma^2+1)^2}\frac{-P^-}{a\lambda_R}, \quad 
       P^{+}(\gamma)\equiv \gamma^{2}+2\gamma -1, \quad P^{-}(\gamma)\equiv P^{+}(-\gamma) ;
            \label{positivequan1}
\end{equation}
the first equation above defines a positive quotient,  restricting  the values of $\gamma$ on the right-hand side;  similarly positivity on the left-hand side of the second equation implies the inequality
\begin{equation}
     \frac{-P^-}{a\lambda_R}> 0.
     \label{ineq4}
\end{equation}
Other relevant quantities for inducing a stable vacuum are  $\det {\cal H}$ for the potentials at the zero energy-point, 
\begin{eqnarray}
     \det {\cal H}_{v_{R}} (v_{0}=0, w_{0}=0)  =  
          \det {\cal H}_{v_{H}} (v_{0}=0, w_{0}=0)= m^4_{R} P^+P^- ;   \label{zerostable2}
\end{eqnarray} 
if $\det {\cal H}>0$, then the zero-energy point will be a minimum or maximum; if  $\det {\cal H}<0$, then it will be a saddle point.
The polynomials of $\gamma$ in Eqs.\ (\ref{positivequan1}), and the polynomial that determines the sign of $\det {\cal H}$ in Eq.\   (\ref{zerostable2}), are shown in the figure (\ref{poly4}).

\begin{figure}[H]
\centering
  \includegraphics[width=.53\textwidth]{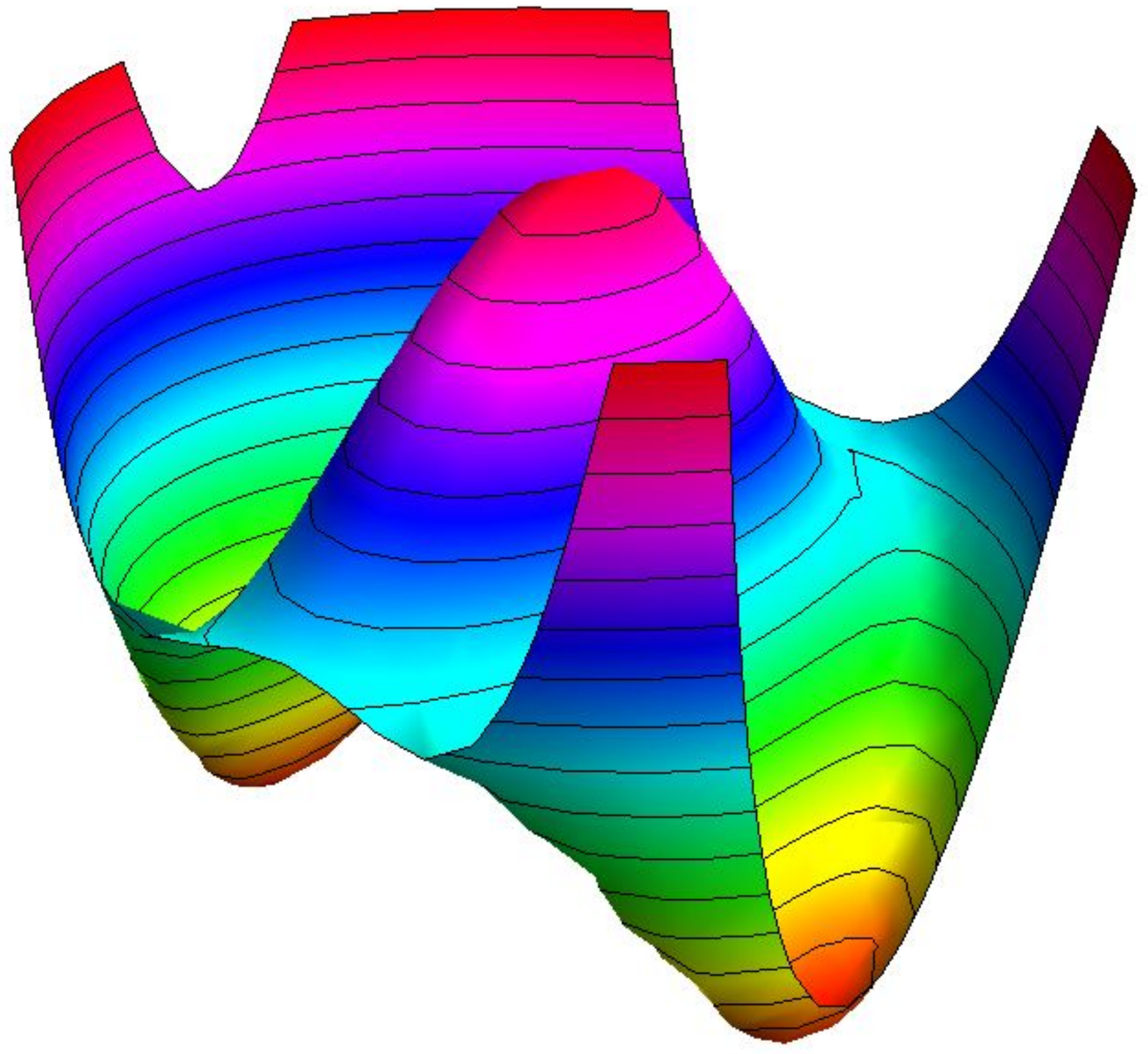}
\caption{ The profile of the potentials $V_{R}/m^2_{R}$, and $V_{H}/m^2_{R}$  in the interval $(1-\sqrt{2},\sqrt{2}-1)$; the zero-energy point corresponds to the  central peak of the potential. The usual $S^1$-valley for the degenerate vacuum is hallowed out at two points, that correspond to the new minima localized at the bottom in the  two red regions, $(\pm v_0,0)$.}
\label{fig11}
\end{figure}

In the figure (\ref{poly4}), the vertical blue asymptotes represent the roots of the polynomial $P^-(\gamma)$, $\gamma=[1-\sqrt{2},1+\sqrt{2}]$; 
hence, the $\lambda$-ratio diverges and $\det {\cal H}$ (red curve) vanishes at such points. Additionally $\det {\cal H}$ vanishes in the roots of
$P^+(\gamma)$, $\gamma=[-1-\sqrt{2},\sqrt{2}-1]$; in the symmetric interval $(1-\sqrt{2},\sqrt{2}-1)$ all polynomials are positive, and a stable vacuum will be induced for the potentials. In this interval, the inequality (\ref{ineq4}) implies that
\begin{equation}
a\lambda_R>0. \label{ineq44}
\end{equation}
Under these simplifications the potentials will take the following form,
\begin{eqnarray}
V_{R}= \frac{a}{2}m^2_{R}(P^{+}v^2+P^{-}w^2)+ \frac{\lambda_{R}}{4!} \Big[(\gamma^2+1)^2\frac{P^+}{P^-}v^{4}+(\gamma^4-8\gamma^3+2\gamma^2+8\gamma+1)\frac{P^+}{P^-}w^4
 +2(\gamma^2+1)^2v^2w^2\Big]; 
     \label{VRSS} \\
V_{H}=\frac{a}{2}m^2_{R}(P^{-}v^2+P^{+}w^2)+ \frac{\lambda_R}{4!} \Big[(\gamma^2+1)^2v^{4}+(\gamma^4+8\gamma^3+2\gamma^2-8\gamma+1)w^4
 +2(\gamma^2+1)^2\frac{P^+}{P^-}v^2w^2\Big]; 
     \label{VHSS}    
     \end{eqnarray} 
the potentials (\ref{VRSS}), and (\ref{VHSS}) are shown in the figure (\ref{fig11}) as functions on $(v,w)$, for a value of $\gamma$ in the interval $(1-\sqrt{2},\sqrt{2}-1)$.
The values $\pm(\sqrt{2}-1)$ are critical, since the expression (\ref{zerostable2}) vanishes, and thus the character of a local maximum for the zero-energy point, and the form of the potentials with stable minima  shown in the figure (\ref{fig11}) are not guaranteed;
therefore one must be careful by taking the limit $\gamma\rightarrow \pm(\sqrt{2}-1)$.

\begin{figure}[H]
  \begin{center}
  \includegraphics[width=.6\textwidth]{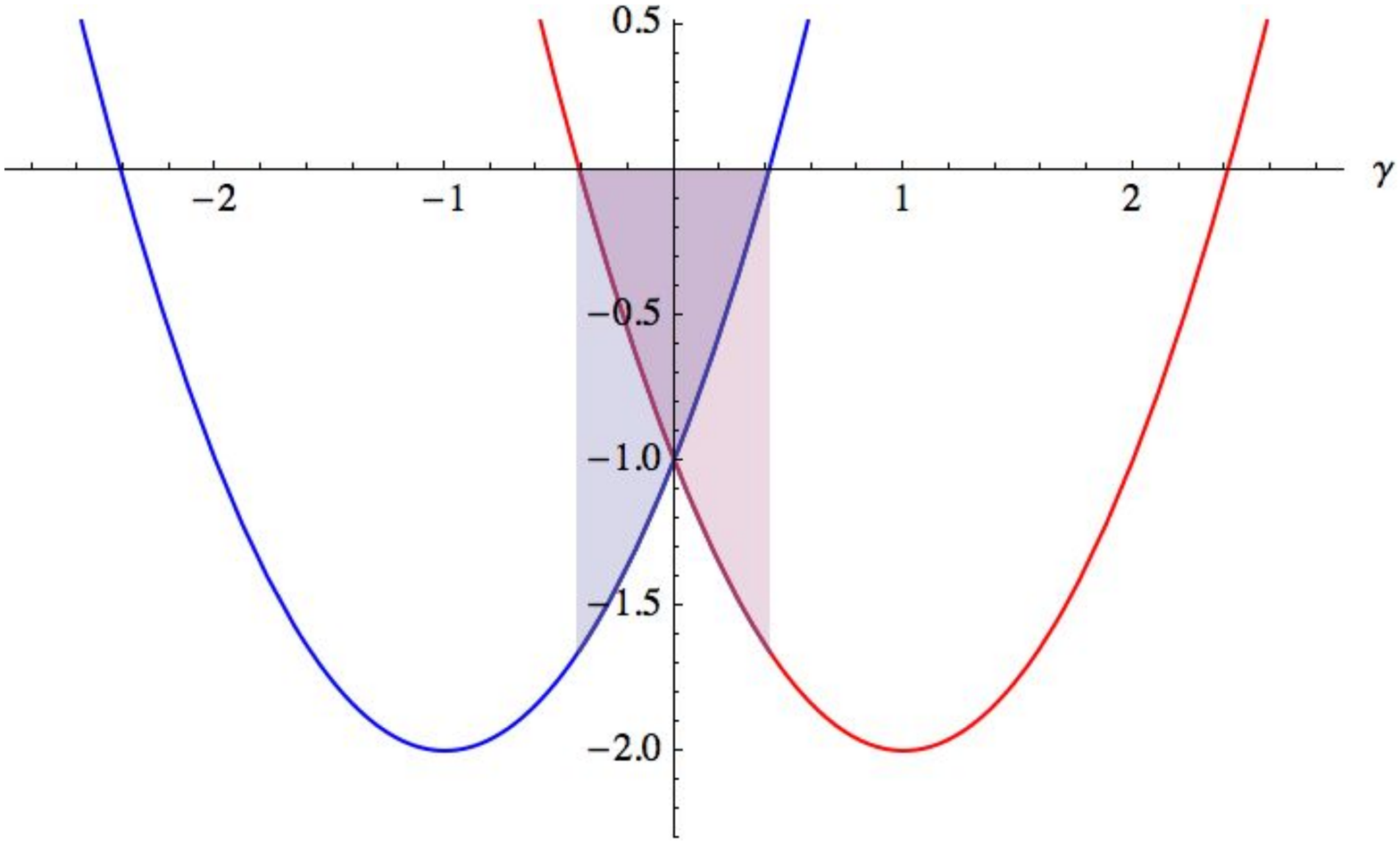}
\end{center}
\caption{The mass polynomial coefficients: the blue curve represents  $P^{+}$, and the red curve represents $P^{-}$. In the shadowed interval $(1-\sqrt{2},\sqrt{2}-1)$ both polynomials are ne\-ga\-ti\-ve, with $P^{+}(\sqrt{2}-1)=0=P^{-}(1-\sqrt{2})$, $P^{+}(1-\sqrt{2})=P^{-}(\sqrt{2}-1)=4(1-\sqrt{2})\approx -1.6569$, and  finally $P^{+}(0)=P^{-}(0)=-1$.}
\label{polmass}
\end{figure}

The polynomials $(P^{+},P^{-})$ that define the mass terms in Eqs. (\ref{VRSS}), and (\ref{VHSS}), are shown in the figure \ref{polmass}; these mass profiles must be compared with the mass polynomial coefficients obtained after the SSB, and described further in the figures \ref{polmassafter}, \ref{kus}, and \ref{kus2}.

The squared expectation value in (\ref{positivequan}) is determined by the polynomial $\frac{-P^{-}}{(\gamma^2+1)^2}$, and is shown in the figure \ref{vevpol}:

\begin{figure}[H]
  \begin{center}
  \includegraphics[width=.6\textwidth]{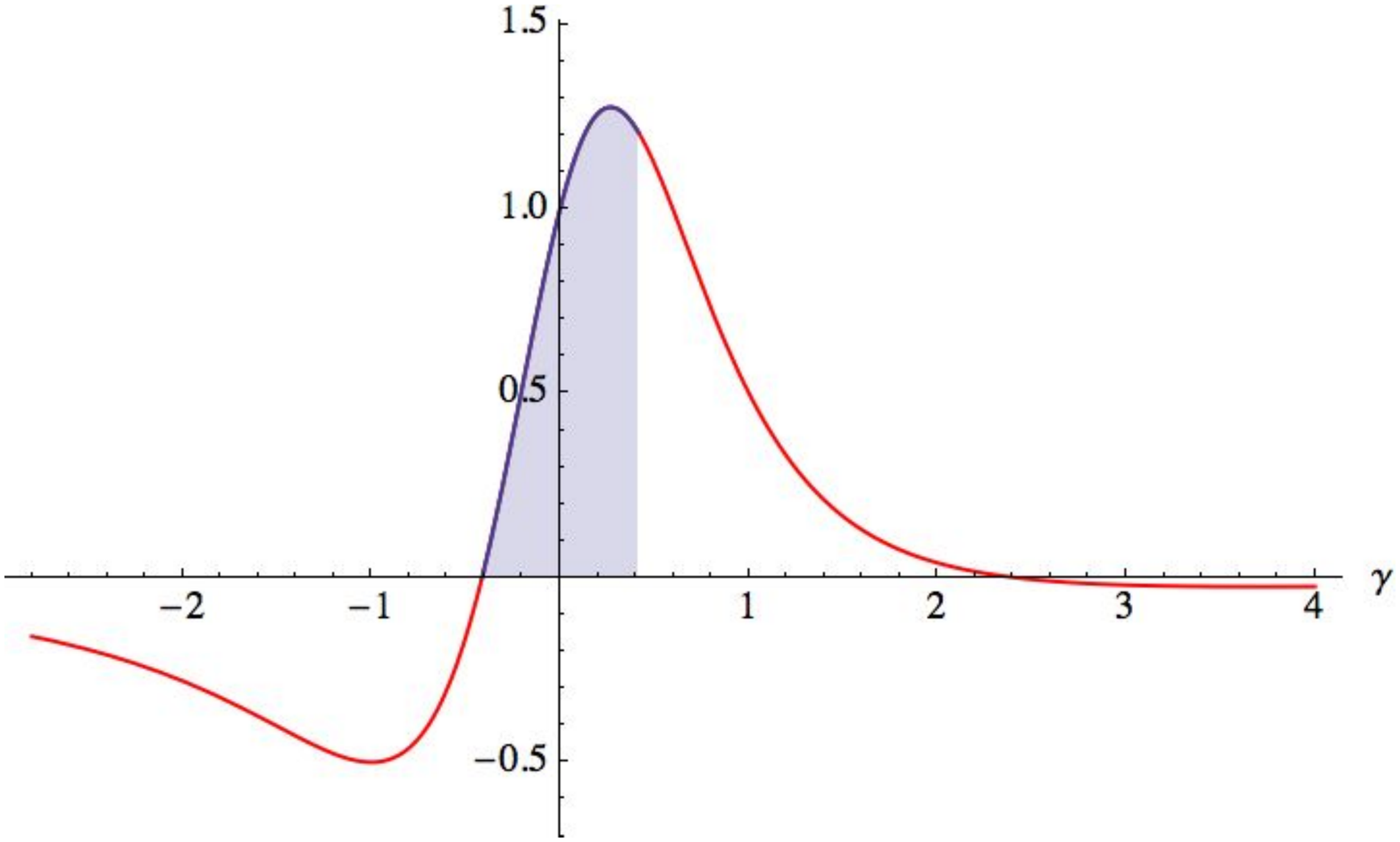}
\end{center}
\caption{ The squared .v.e.v. as a $\gamma$-deformation of the usual value $\frac{6m^2_{R}}{a\lambda_{R}}$; we have that $\frac{-P^{-}}{(\gamma^2+1)^2}\Big|_{\gamma=0}=1$, and at the limits of the shadowed interval, $\frac{-P^{-}}{(\gamma^2+1)^2}\Big|_{\gamma=1-\sqrt{2}}=0$, and  $\frac{-P^{-}}{(\gamma^2+1)^2}\Big|_{\gamma=\sqrt{2}-1}=\frac{1}{2} + \frac{1}{\sqrt{2}}\approx 1.2071$. Within this interval, the polynomial has a maximum value at
$\gamma=2-\sqrt{3}\approx 0.2679$.}
\label{vevpol}
\end{figure}
Furthermore, the vacuum energies are given by
\begin{eqnarray}
V_R(\pm v_0, 0)= \frac{-3m^4_R}{2\lambda_R}\frac{P^+P^-}{(\gamma^2+1)^2},\qquad V_H(\pm v_0, 0)= \frac{-3m^4_R}{2\lambda_R}\frac{(P^-)^2}{(\gamma^2+1)^2};
\label{vacenergy}
\end{eqnarray}
hence, the depth of the red regions in the figure \ref{fig11} depends on $\gamma$; the polynomials  that deform the conventional vacuum energies in the above expressions are shown in the figure \ref{vacenergytwo}. For $\gamma=0$ one can recover 
from $V_{R}$ or $V_{H}$ the vacuum energy for the usual $U(1)$ field theory, since  $\frac{P^+P^-}{(\gamma^2+1)^2}\Big|_{\gamma=0}=1=\frac{(P^-)^2}{(\gamma^2+1)^2}\Big|_{\gamma=0}$. The 
 polynomials take values in the interval $(0,1)$ for $V_{R}$, and in the interval $(0,2)$ for $V_{H}$.
 \begin{figure}[H]
  \includegraphics[width=.5\textwidth]{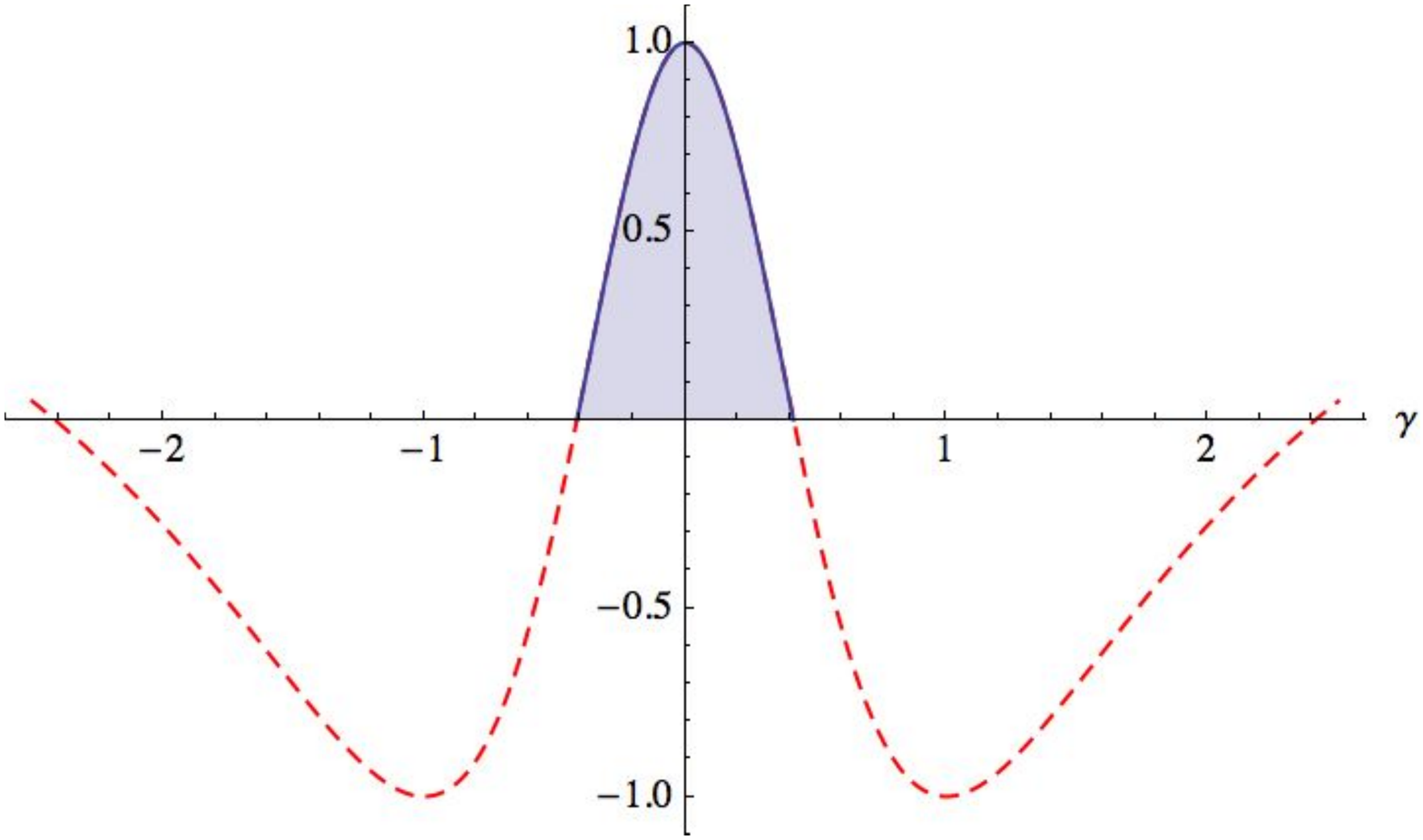}
  \includegraphics[width=.5\textwidth]{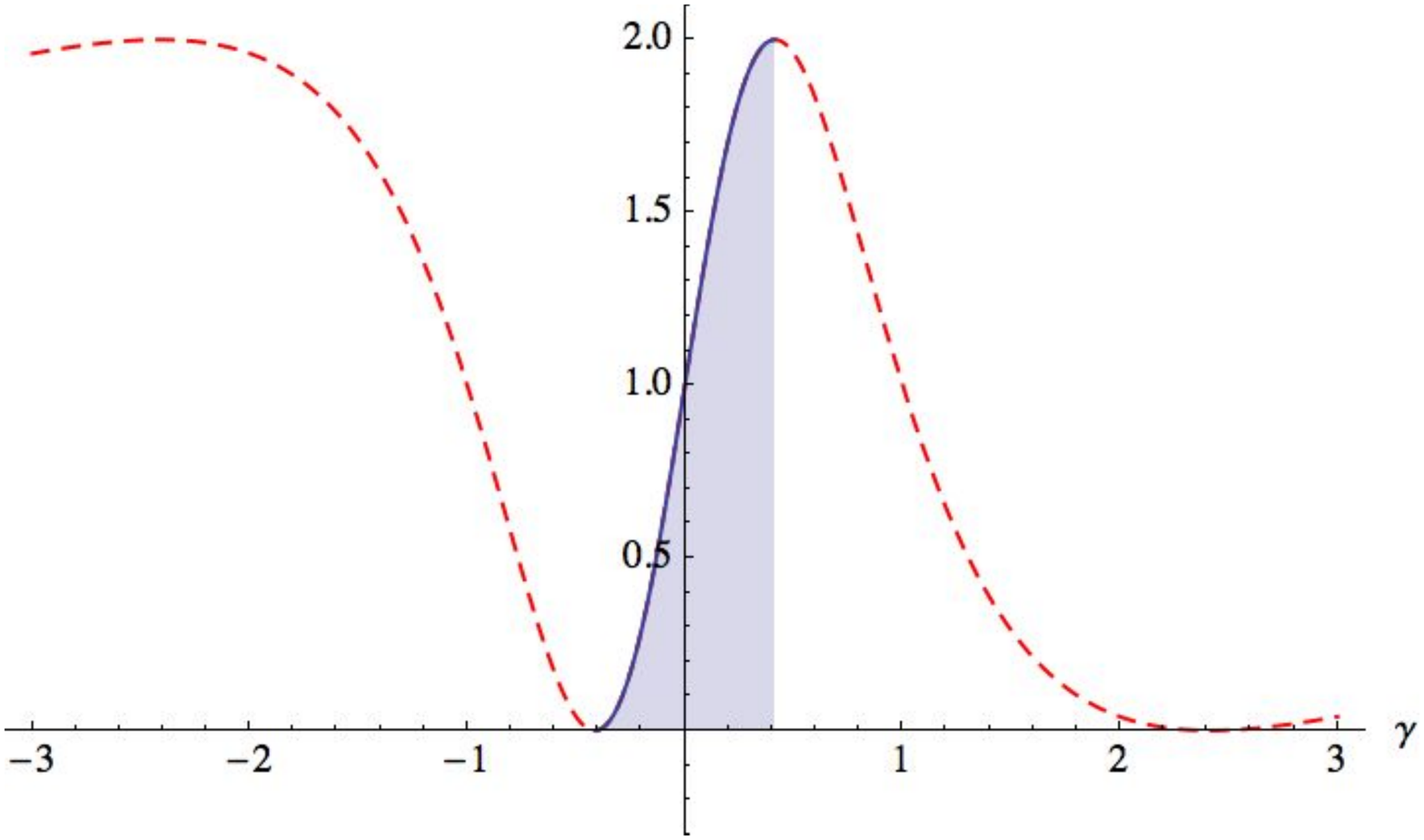}
\caption{On the left panel the polynomial  $\frac{P^+P^-}{(\gamma^2+1)^2}$, and on the right panel $\frac{(P^-)^2}{(\gamma^2+1)^2}$. In the interval $(1-\sqrt{2},\sqrt{2}-1)$ the $\gamma$-deformations take positive values, and the vacuum energies are finite, even at the limits of the interval.}
\label{vacenergytwo}
\end{figure}
We now describe the effective interaction terms;
\begin{figure}[H]
  \begin{center}
  \includegraphics[width=.6\textwidth]{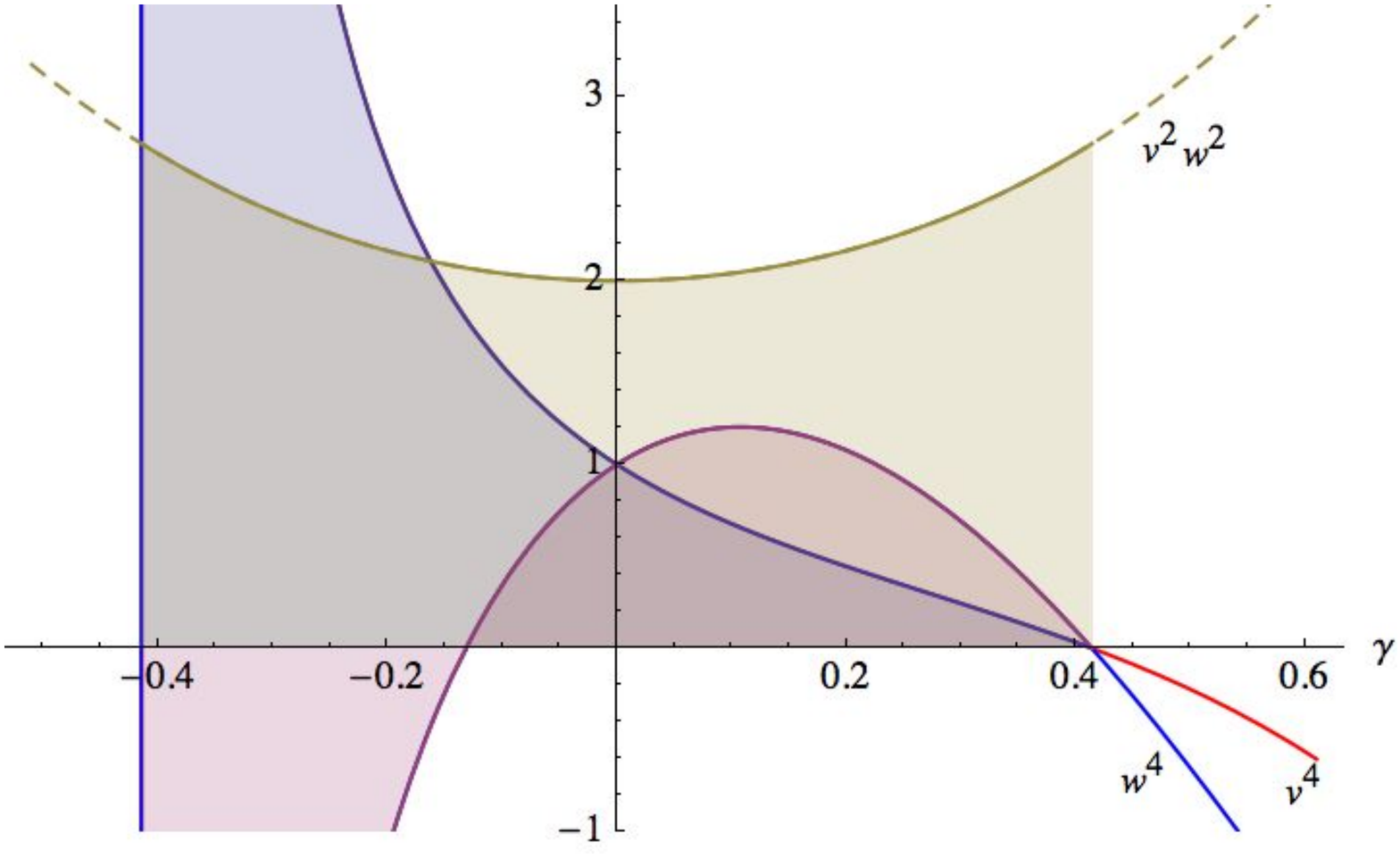}
\end{center}
\caption{ The self-interactions terms for $v$, and $w$, and the cross-coupling term in the potential $V_R$ in Eq. (\ref{VRSS}).}
\label{intt1}
\end{figure}
In the figure \ref{intt1} we see that at the right limit $\sqrt{2}-1$,  the fields are weakly self-interacting, and strongly coupling to each other;
furthermore, at the left limit $1-\sqrt{2}$, the cross coupling is the same, but the self-interactions diverge, and in particular the self-interaction for
$w$ takes negative values. The roots for the polynomial $\gamma^4-8\gamma^3+2\gamma^2+8\gamma+1$ that appears in the self-interaction term for $w$ are,
\begin{eqnarray}
2 - \sqrt{6} + \sqrt{5 - 2 \sqrt{6}}\approx -0.13165, \quad 2 - \sqrt{6} -\sqrt{5 - 2 \sqrt{6}}\approx -0.76733, \nonumber\\
2 + \sqrt{6} +\sqrt{5 +2 \sqrt{6}}\approx 7.5958, \quad 2 +\sqrt{6} - \sqrt{5 +2 \sqrt{6}}\approx 1.3032;
\label{rootsw}
\end{eqnarray}
the first one is shown in the figure \ref{intt1}, and thus restricts the interval by the left hand side.
For the potential $V_H$ we have,
\begin{figure}[H]
  \begin{center}
  \includegraphics[width=.6\textwidth]{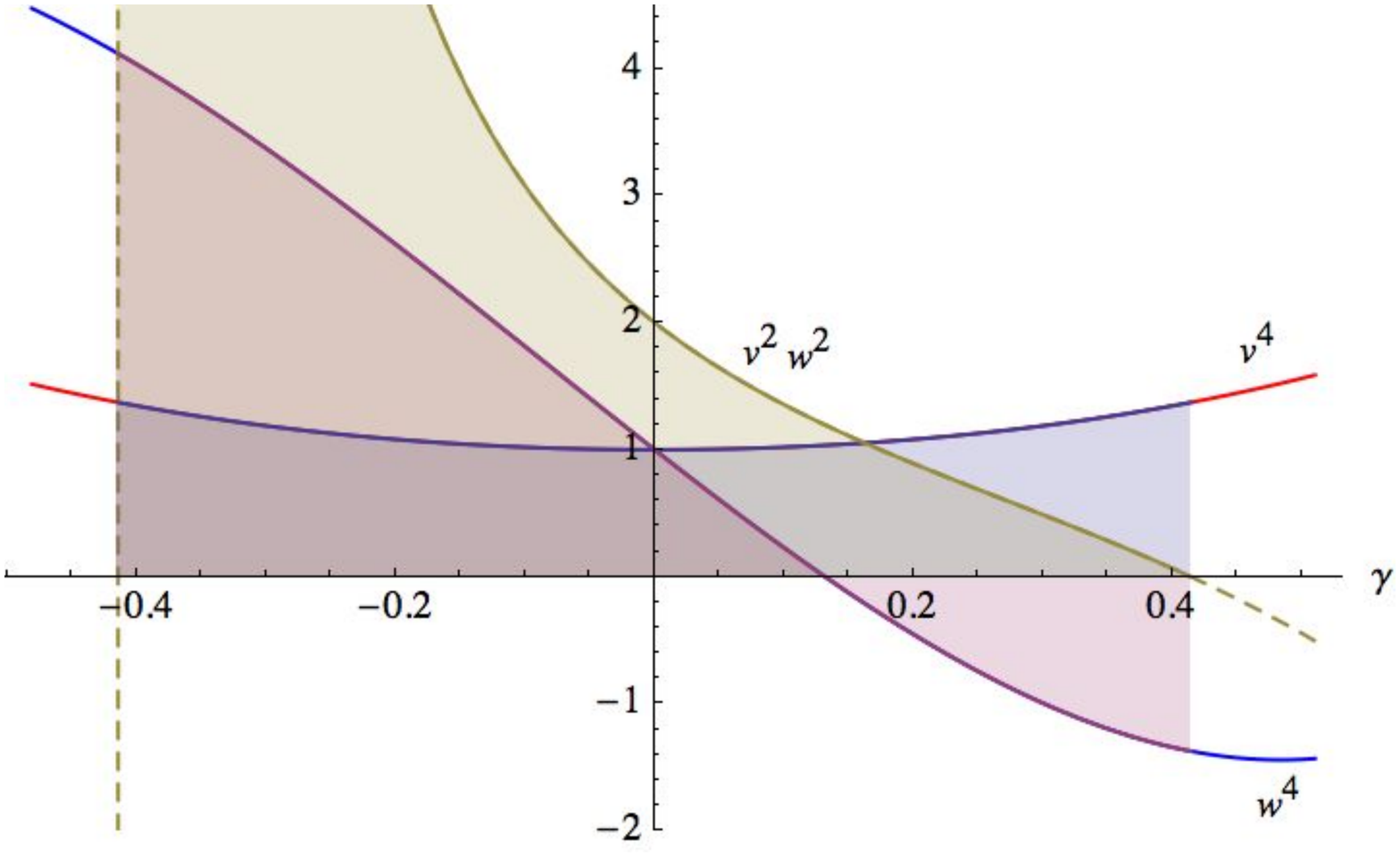}
\end{center}
\caption{ The self-interactions terms for $v$, and $w$, and the cross-coupling term in the potential $V_H$ in Eq. (\ref{VHSS}).}
\label{intt2}
\end{figure}
Additionally in the figure \ref{intt2}, at the right limit $\sqrt{2}-1$,  the fields are weakly interacting; the field $v$ is strongly self-interacting; however at this limit, 
the effective self-coupling for $w$ takes negative values, and one must restrict the original range by the right hand side too; the new right limit will be determined by a root of the effective polynomial for $w$.
Furthermore, at the left limit $1-\sqrt{2}$, the cross coupling diverges, with finite self-interactions; hence there is no a duality between the potentials $V_R$, and $V_H$. The divergences at this limit are avoided by the restriction of the left hand side determined by the root in Eq. (\ref{rootsw}).

The polynomial that defines the self-interaction of $w$ in $V_H$, $\gamma^4+8\gamma^3+2\gamma^2-8\gamma+1$, is the mirror polynomial of that considered in Eq. (\ref{rootsw}), with four roots that correspond, with a change of sign, to those shown in Eq. (\ref{rootsw}); in particular the root shown in the figure \ref{intt2} is just a mirror root of that shown in the figure \ref{intt1}; thus, the new allowed range is the restricted symmetric range,
 \begin{equation}
 \Big[2 - \sqrt{6} + \sqrt{5 - 2 \sqrt{6}},-2 + \sqrt{6} - \sqrt{5 - 2 \sqrt{6}}\Big],
  \label{newran}
   \end{equation}
with the limits included.

\subsection{Usual gauge}
\label{usualgg}
\begin{figure}[H]
  \begin{center}
  \includegraphics[width=.6\textwidth]{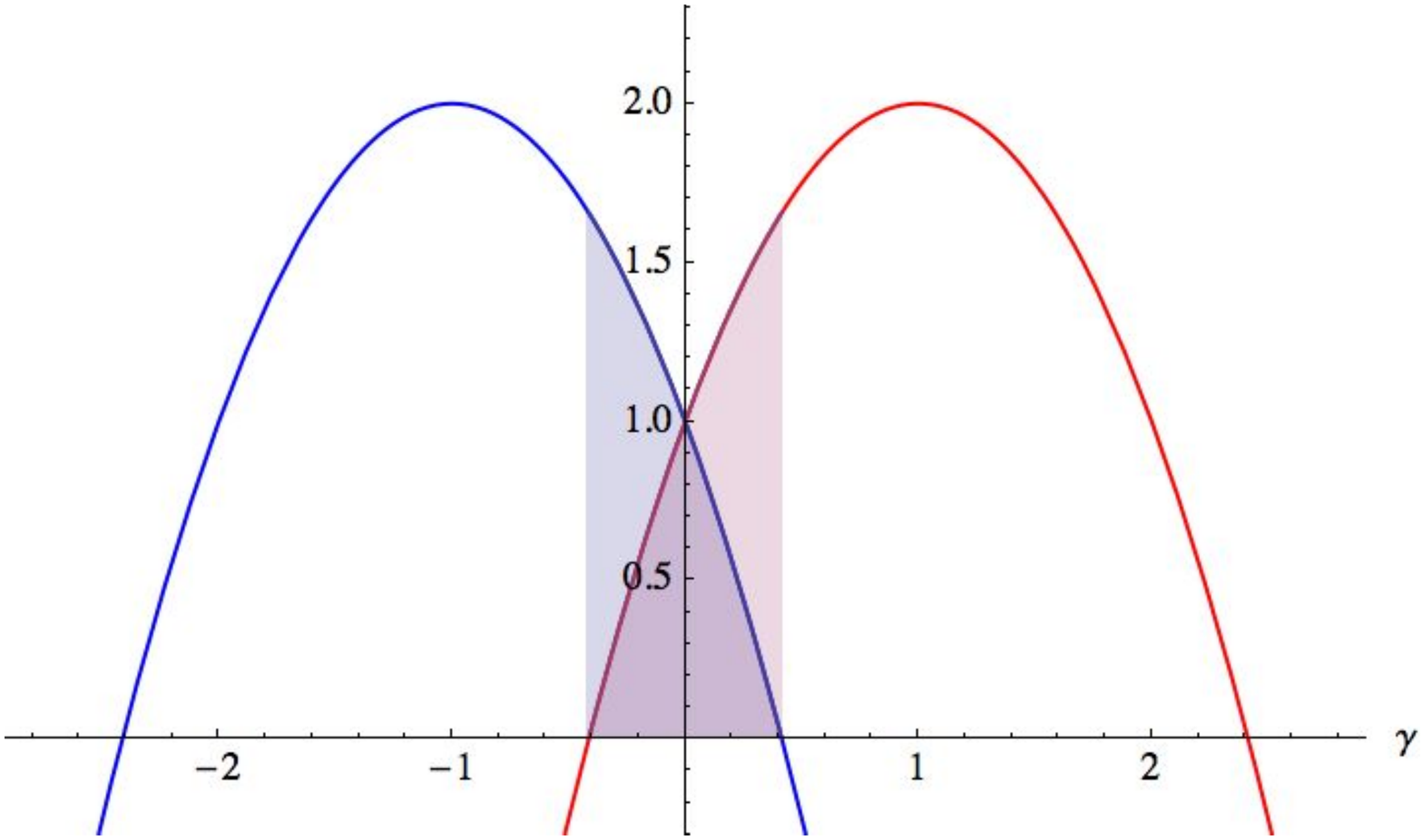}
\end{center}
\caption{ In blue the running mass coefficient $-P^+ (\gamma)$ after SSB, and in red $-P^- (\gamma)$; the case $\gamma=0$ reproduces the usual SSB of $U(1)$. The field $w$ is a fully massless for any $\gamma$ in the interval.}
\label{polmassafter}
\end{figure}

The circular and hyperbolic rotations can be spontaneously broken by the choice (\ref{usualchoice}), and  remembering that in this case we are considering 
  the vacuum expectation values  $(v_0\neq 0, w_0=0)$ given in  (\ref{positivequan1}); thus all mixed terms $vw$ (both real and hybrid) can be gauged away, reducing the quadratic terms in Eq.\ (\ref{quacub1}) to the canonical form,
\begin{eqnarray}
      -\frac{am^2}{2}\psi\overline{\psi} + \frac{\lambda}{4!}(\overline{\psi}^{2}_{0} \psi^{2} + \psi^{2}_{0} \overline{\psi}^{2} ) 
= am^2_{R}(-P^+ - kP^-)v^2;
\label{ssb4} 
\end{eqnarray} 
where we have used the $\lambda$-ratio (\ref{positivequan1}); the field $w$ is massless in both senses, real and hybrid; furthermore, the field $v$ has duplicated its real and $k$-hybrid masses with a change of sign. The figure \ref{polmassafter} shows that 
 when the real mass goes to zero as $\gamma\rightarrow\sqrt{2}-1$, the hybrid mass goes to its maximum value, and in reverse in the limit $\gamma\rightarrow 1-\sqrt{2}$; the masses coincide in the value $m^2_R$ for $\gamma=0$.
Since that the field $w$ is massless in both senses, there is no mass hierarchy between the masses for different fields; however one can establish a hierarchy between the real and the $k$-mass for the field $v$  generated in (\ref{ssb4}); the criterion for fixing the $\gamma$-parameter may be the critical value for the maximum value of the v.e.v. described in the figure \ref{vevpol},
\begin{equation}
-P^+(\gamma\approx 0.2679)\approx  0.39243, \quad -P^-(\gamma\approx 0.2679) \approx  1.46403;
\label{vev}
\end{equation}
with a hierarchy of order $10$ in favor of the $k$-mass; however, this critical value is out of the restricted range (\ref{newran}). Thus we can use the limits of such a range for adjusting the hierarchy shown in Eq. (\ref{vev});  at these limits the field $w$ is not self-interacting;
\begin{eqnarray}
-P^+(\gamma\approx 0.13165)\approx  0.71936, \quad -P^-(\gamma\approx0.13165) \approx  1.2460; \nonumber\\ 
-P^+(\gamma\approx -0.13165)\approx  1.2460, \quad -P^-(\gamma\approx -0.13165) \approx   0.71936.
\label{noselfw}
\end{eqnarray}
This case although simple, will allow us to compare the mass spectrum generated through the usual gauge (\ref{usualchoice}), to that generated by the gauge (\ref{diagonal4}).

\subsection{Circular and hyperbolic parameters in terms of the $\gamma$-parameter}
\label{usinggamma}
The gauge (\ref{diagonal4}), and the  expressions (\ref{positivequan1}), lead to the following canonical form for the mass terms,
\begin{equation}
      -\frac{am^2}{2}\psi\overline{\psi} + \frac{\lambda}{4!}(\overline{\psi}^{2}_{0} \psi^{2} + \psi^{2}_{0} \overline{\psi}^{2} ) 
= am^2_{R}\Big(Q^v_+(\gamma) v^2+Q^w_+(\gamma) w^2+kQ^v_-(\gamma)v^2+kQ^w_-(\gamma)w^2\Big);
\label{prime} 
\end{equation} 
where
\begin{eqnarray}
Q^v_+(\gamma;l,s)\equiv-\frac{1}{2}P^+ -\frac{1}{2}\frac{P^+P^-[s(\gamma^2-1)^3P^--8l\gamma^3P^+]+4\gamma(\gamma^2-1)[s(\gamma^2-1)^3P^++8l\gamma^3P^-]}{(\gamma^2+1)^4\sqrt{P^+P^-}},\label{q1}\\
Q^w_+(\gamma;l,s)\equiv -\frac{1}{2}P^--\frac{1}{2}\frac{-s(\gamma^2-1)^3P^-+8l\gamma^3P^+}{(\gamma^2+1)^2\sqrt{P^+P^-}}, \label{q2}\\
Q^v_-(\gamma;l,s)\equiv Q^v_+(-\gamma;l,s),\quad Q^w_-(\gamma;l,s)\equiv Q^w_+(-\gamma;l,s);\label{q3mirror}
\end{eqnarray}
and $l=\pm 1$, $s=\pm 1$, with $ls=-1$. The mirror polynomials $Q^v_-$, and $Q^w_-$ are obtained from $Q^v_+$, and $Q^w_+$ by the transformation
$\gamma\rightarrow -\gamma$, and determine the corresponding hybrid $k$-mass terms.

As opposed to the expression (\ref{ssb4}), the expression (\ref{prime}) gives the possibility that the field $w$ may be massive in both senses.
Furthermore, the figure \ref{polmassafter} shows that there is no value for $\gamma$ for which the field $v$ is massless in both senses under the usual gauge; the case at hand will offer such a possibility, although with a massive field $w$.
Note also that in the expression (\ref{pprime}), the $k$-mass terms are not determined by mirror polynomials, such as the above equations. We consider now the choices for the pair $(l,s)$.

\subsubsection{$s=1$, and $l=-1$}
\begin{figure}[H]
  \begin{center}
  \includegraphics[width=.6\textwidth]{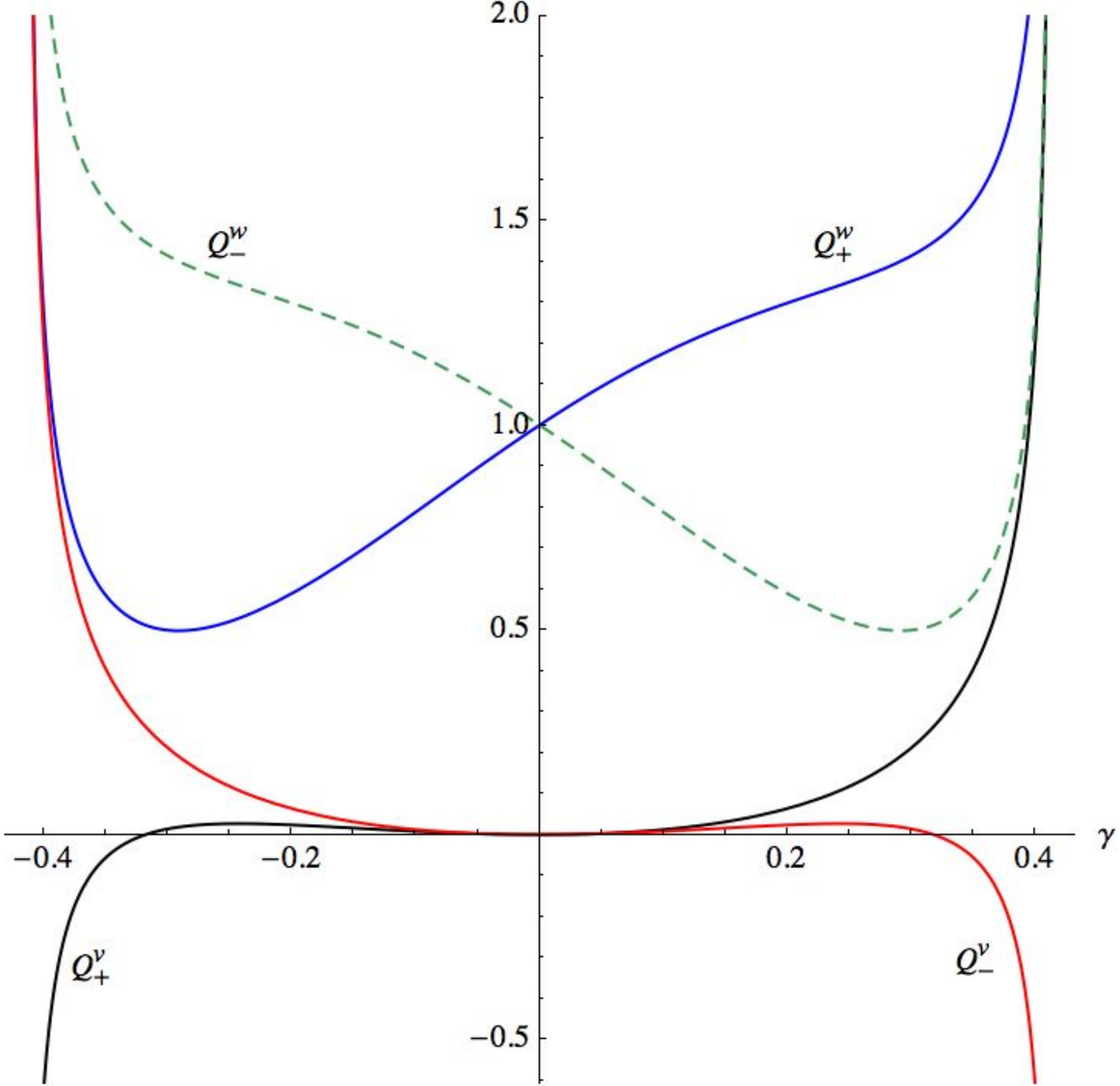}
\end{center}
\caption{ the polynomials $Q^v_+$ (black curve), $Q^w_+$ (blue curve), and their mirror polynomials  $Q^v_-$ (red curve), $Q^w_-$ (dashed curve)
 for $s=1$, and $l=-1$, in the interval $\gamma \in (1-\sqrt{2},\sqrt{2}-1)$; in the limits of this interval the polynomials diverge.}
\label{kus}
\end{figure}
In figure \ref{kus} we see that the field $v$ is massless for $\gamma=0$, in both senses, real and $k$-hybrid, $Q^v_+(\gamma=0)=0=Q^v_-(\gamma=0)$; and for the field $w$ we have that $Q^w_+(\gamma=0)=1=Q^w_-(\gamma=0)$. This result can be considered as exotic, since the field $v$ has a non-zero vacuum expectation value, and can result massless after the spontaneous symmetry breaking, and conversely in relation to the field $w$, since it has a zero vacuum expectation value, and can acquire mass in both senses, real and hybrid.

Furthermore, there are two roots $\pm \sqrt{5 - 2 \sqrt{6}}\approx \pm 0.3178$ where the field $v$ is semi-massless, with a pure real mass, or pure hybrid mass depending on the sign chosen.

{\bf A root of $Q^v_+$:}
\begin{eqnarray}
Q^v_+\Big(-\sqrt{5 - 2 \sqrt{6}}\Big)=0; \quad Q^v_-\Big(-\sqrt{5 - 2 \sqrt{6}}\Big)\approx 0.2633;\nonumber\\
Q^w_-\Big(-\sqrt{5 - 2 \sqrt{6}}\Big)\approx 1.4469,\quad Q^w_+\Big(-\sqrt{5 - 2 \sqrt{6}}\Big)\approx  0.5116;
\label{semimass}
\end{eqnarray}
in this case the hierarchy is in favor of the $k$-mass for the field $w$, with a order of $10$ with respect to its real mass, and with respect to the $k$-mass of the field $v$; and similarly for the other root,  $\sqrt{5 - 2 \sqrt{6}}$. However, both roots are out of the restricted range (\ref{newran}).
In the interval $\gamma \in (1-\sqrt{2},\sqrt{2}-1)$ (and in the restricted range), the polynomials $Q^w_+(\gamma)$, and $Q^w_-(\gamma)$ are strictly positive; $Q^w_+(\gamma)$ has its minimum value at $\gamma\approx -0.2912$, a close value to the root described above.

{\bf Global minimum for $Q^w_+(\gamma)$:}
\begin{eqnarray}
Q^w_+(-0.2912)\approx 0.4979 \quad Q^w_-(-0.2912)\approx 1.4001;\nonumber\\
Q^v_-(-0.2912)\approx 0.1909 ; \quad Q^v_+(- 0.2912)\approx   0.0183;
\label{mini}
\end{eqnarray}
in this case the hierarchy is basically that described just above in favor of the $k$-mass for $w$, but additionally a hierarchy of order $10^2$
appears in relation to the real mass of the field $v$. This critical point is also out of the restricted range. Therefore, the last critical points correpond just to the limits of the restricted range:

{\bf Vanishing self-interaction for field $w$: right limit}
\begin{eqnarray}
Q^w_+(-2 + \sqrt{6} - \sqrt{5 - 2 \sqrt{6}})\approx 1.2196 \quad Q^w_-(-2 + \sqrt{6} - \sqrt{5 - 2 \sqrt{6}})\approx .71961;\nonumber\\
Q^v_-(-2 + \sqrt{6} - \sqrt{5 - 2 \sqrt{6}})\approx 1.2969\times 10^{-2} ; \quad Q^v_+(-2 + \sqrt{6} - \sqrt{5 - 2 \sqrt{6}})\approx  2.2296 \times 10^{-2};
\label{newmini}
\end{eqnarray}
hence, we have comparable masses for the same field, and a hierarchy of order $10^{2}$ in favor of the field $w$; this result may be considered as exotic, since the field $w$ has a vanishing v.e.v. For the left limit, the hierarchy is essentially the same, with an interchange between the labels $(+\leftrightarrow-)$.

\subsubsection{the case $s=-1$, and $l=1$}
\begin{figure}[H]
  \begin{center}
  \includegraphics[width=.6\textwidth]{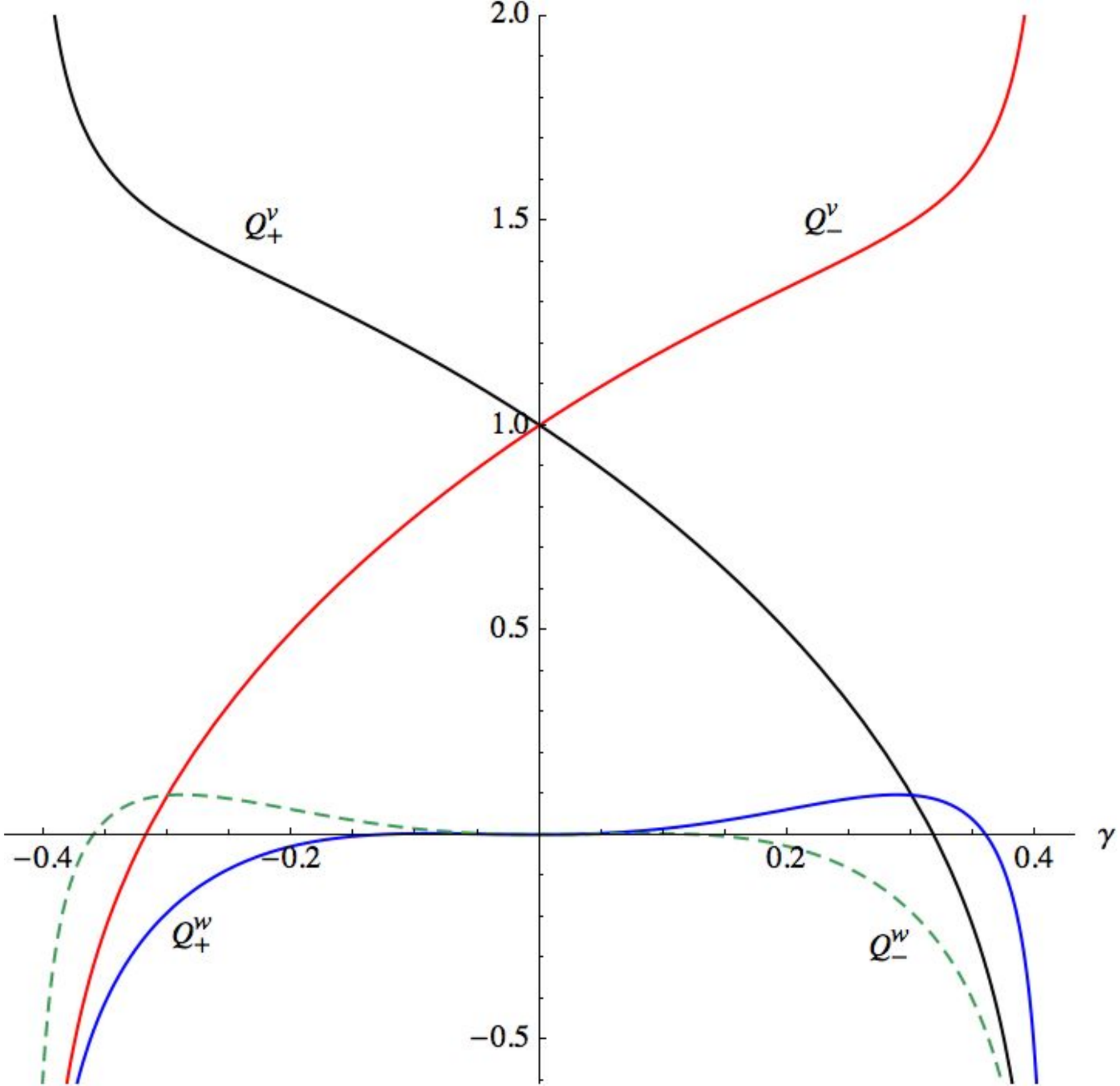}
\end{center}
\caption{ the polynomials $Q^v_+(\gamma)$ (black curve), $Q^w_+(\gamma)$ (blue curve), and their mirror polynomials  $Q^v_-(\gamma)$ (red curve), $Q^w_-(\gamma)$ (dashed curve)
 for $s=-1$, and $l=1$, in the interval $\gamma \in (1-\sqrt{2},\sqrt{2}-1)$; in the limits of this interval the polynomials diverge.}
\label{kus2}
\end{figure}
In the figure (\ref{kus2}) the roots $\pm \sqrt{5 - 2 \sqrt{6}}$ appear again,

{\bf A root of $Q^v_+$:}
 \begin{eqnarray}
Q^v_+\Big(\sqrt{5 - 2 \sqrt{6}}\Big)=0; \quad Q^v_-\Big(\sqrt{5 - 2 \sqrt{6}}\Big)\approx 1.5347; \nonumber\\
Q^w_+\Big(\sqrt{5 - 2 \sqrt{6}}\Big)\approx 0.0877,\quad Q^w_-\Big(\sqrt{5 - 2 \sqrt{6}}\Big)\approx  -0.2482;
\label{semimass2}
\end{eqnarray}
furthermore, in this case $Q^w_+$ and its mirror polynomial have roots, as opposed to the case discussed in the figure  (\ref{kus}),

{\bf A root of $Q^w_+$:}
\begin{eqnarray}
Q^w_+(0.3599)=0, \quad Q^w_-(0.3599) \approx -0.4797;\nonumber\\
Q^v_+(0.3599)\approx -0.3207, \quad Q^v_-(0.3599)\approx 1.6803.
\label{QQ}
\end{eqnarray}
On the other hand, $Q^w_+$ ( its mirror polynomial) has a global maximum at $\gamma \approx 0.2887$ ($\gamma \approx -0.2887$);

{\bf A global maximum for $Q^w_+$:}
\begin{eqnarray}
Q^w_+(0.2887)\approx 0.0975, \quad Q^w_-(0.2887) \approx -0.1588;\nonumber\\
Q^v_+(0.2887)\approx 0.1538, \quad Q^v_-(0.2887)\approx 1.4747.
\label{Q3}
\end{eqnarray}
The crossing point for the black and blue curves is at $\gamma\approx0.3003$;

{\bf A crossing point of $Q^w_+$, and $Q^v_+$:}
\begin{eqnarray}
Q^w_+(0.3003)\approx 0.0961 \approx Q^v_+(0.3003), \quad Q^v_-(0.3003) \approx 1.4968 \quad Q^w_-(0.3003)\approx -0.1901;
\label{Q4}
\end{eqnarray}
similarly the crossing point for the dashed and red curves is at $\gamma\approx-0.3003$. 
However, in the expressions (\ref{semimass2}), (\ref{QQ}), (\ref{Q3}) and (\ref{Q4}), some polynomials take negative values, and all critical $\gamma$-values lie out of the restricted range (\ref{newran}); hence,
 we do not consider such cases as viable. Now, a zoom of the figure  \ref{kus2} will show new critical points for $\gamma$, with admissible values for $\gamma$ within the restricted range.
\begin{figure}[H]
 \begin{center}
  \includegraphics[width=.6\textwidth]{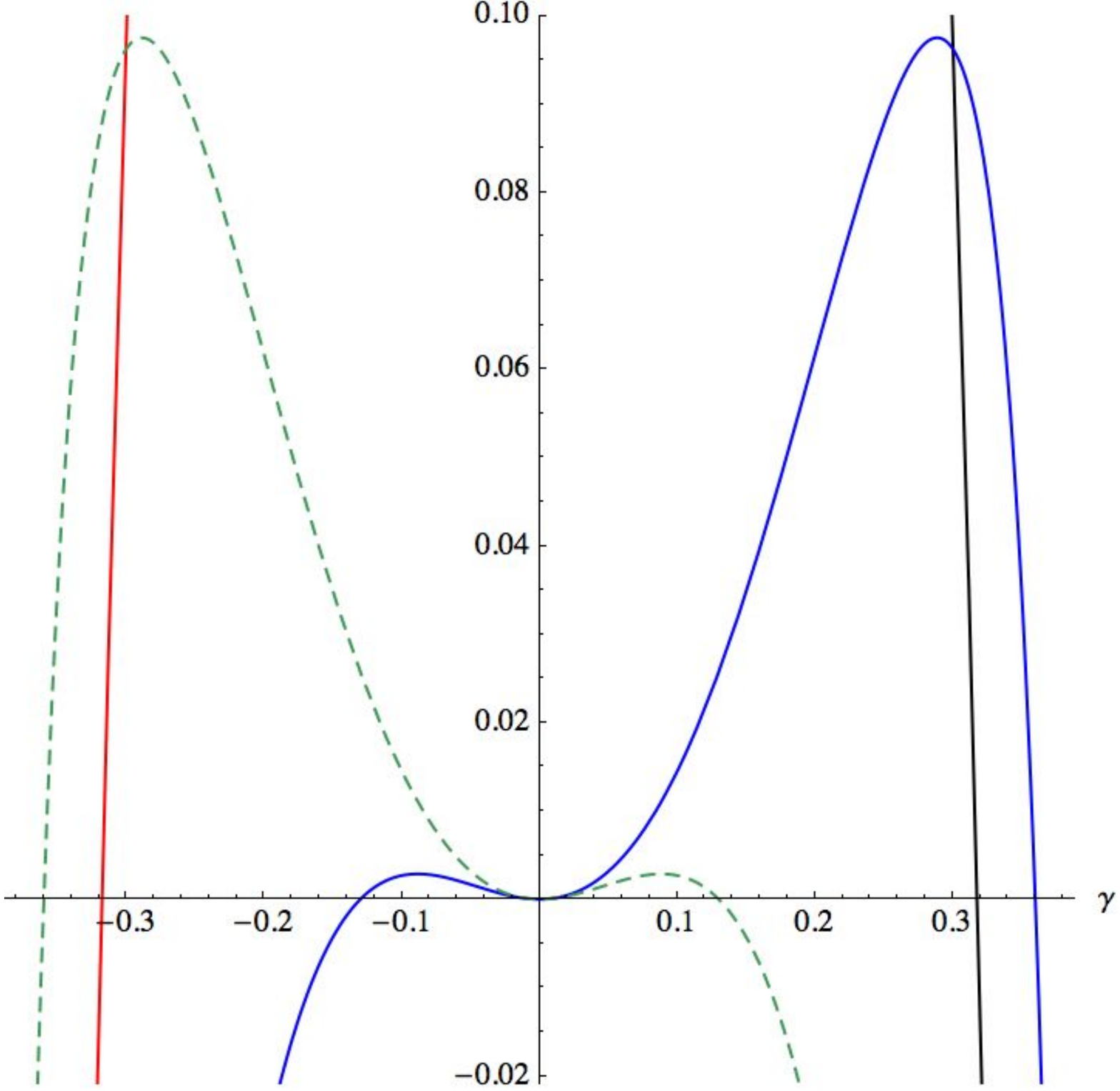}
\end{center}
\caption{A zoom of the figure (\ref{kus2});  new roots at $\gamma \approx \pm 0.1301$, and local maxima at $\gamma \approx \pm 0.0884$, are shown for $Q^w_+$ and its mirror polynomial.}
\label{kuszoom}
\end{figure}
In the figure (\ref{kuszoom}), we have at the root of $Q^w_+$,

{\bf A root of  $Q^w_+$:}
\begin{eqnarray}
Q^w_+(-0.1301)= 0, \quad Q^w_-(-0.1301)\approx 0.0257,\quad Q^v_-(-0.1301)\approx 0.7005, \quad  Q^v_+(-0.1301)\approx1.2305;
\label{kuskus}
\end{eqnarray}
where the masses of $v$ are comparable, and with a hierarchy of order $10^2$ in relation to the $k$-mass of $w$. At this root, 
the interactions  polynomials in the potential $V_R$ take the approximate values,
\begin{eqnarray}
v^4: 1.7786, \qquad w^4: 1.8841\times 10^{-2},\qquad v^2w^2: 2.0682;
\label{rootint}
\end{eqnarray}
where the relative smallness of the polynomial of $w$ is due to the closeness of the root of $Q^w_+$, and the root of the self-interaction polynomial of $w$. Similarly, for the potential $V_H$ we have that,
\begin{eqnarray}
v^4: 1.03414, \qquad w^4: 2.05732,\qquad v^2w^2: 3.5572;
\label{rootint2}
\end{eqnarray}
these values are of the same order, in contrast with those in the Eq. (\ref{rootint}).

Furthermore, at the local maximum for $Q^w_+$,
 
{\bf A  local maximum for $Q^w_+$:}
\begin{eqnarray}
Q^w_+(-0.0884)\approx 0.0028, \qquad\qquad Q^w_-(-0.0884)\approx 0.0110,  \\
Q^v_-(-0.0884)\approx 0.8060, \qquad\qquad  Q^v_+(-0.0884)\approx 1.1624;
\label{kuskus2}
\end{eqnarray}
the hierarchy is basically the same described in the expressions (\ref{kuskus}), but with an additional hierarchy of order $10^3$ in relation to the real mass of $w$. For the interactions we have,
\begin{eqnarray}
V_R:  \qquad v^4: 1.4561, \qquad w^4: 0.4502,\qquad v^2w^2: 2.0313; \label{vrint}\\
V_H:  \qquad v^4: 1.01569, \qquad w^4: 1.7173,\qquad v^2w^2: 2.9123.
\label{vhint}
\end{eqnarray}

\section{The case $(v_0= 0,w_0\neq 0)$, and $m_{R}=m_{H}$.}
\label{nogamma11}

Along the same lines followed in Section \ref{gammano1}, the use of the Eq. (\ref{expvalue2}) leads
 essentially to the same expressions (\ref{positivequan1}), (\ref{ineq4}), and (\ref{zerostable2}), with the change $\gamma\rightarrow-\gamma$; 
 \begin{equation}
     \frac{\lambda_{H}}{\lambda_{R}} =\frac{P^-}{P^+},    
       \quad w^{2}_{0} = \frac{6m^{2}_{R}}{(\gamma^2+1)^2}\frac{-P^+}{a\lambda_R};
            \label{cocientlambda}
\end{equation}
 in the figure (\ref{poly4}) the red curve is the same for this case, since $\det{\cal H}$ is invariant under $\gamma\rightarrow-\gamma$. However the blue and dashed curves will be mirrored on the $``y"$ axis due to the change $\gamma\rightarrow-\gamma$. Positivity is maintained in the range $(1-\sqrt{2},\sqrt{2}-1)$, hence the inequality (\ref{ineq44}) remains valid; the potentials can be described in terms of the same mass polynomials appearing in the expressions (\ref{VRSS}) and (\ref{VHSS}), but with the interchange $ P^+\leftrightarrow P^-$ in the $\lambda_{R}$-terms,
\begin{eqnarray}
V_{R}=  \frac{a}{2}m^2_{R}(P^{+}v^2+P^{-}w^2)+ \frac{\lambda_{R}}{6} \Big[ P^+\leftrightarrow P^-\Big]; 
     \label{VRSSS} \\
V_{H}=\frac{a}{2}m^2_{R}(P^{-}v^2+P^{+}w^2)+ \frac{\lambda_{R}}{6} \Big[  P^+\leftrightarrow P^- \Big]; 
     \label{VHSSS}    
     \end{eqnarray} 
these potentials have the same form shown in the figure (\ref{fig11}). In this case
 the field $v$ will be massless, and the field $w$ will develop a mass with both parts, real and $k$-hybrid; under the choice (\ref{usualchoice}) we have that,
\begin{eqnarray}
      -\frac{am^2}{2}\psi\overline{\psi} + \frac{\lambda}{4!}(\overline{\psi}^{2}_{0} \psi^{2} + \psi^{2}_{0} \overline{\psi}^{2} ) = -am^2_{R}(P^- + k P^+)w^2;\label{doubleSS}
\end{eqnarray}
which can be obtained from the expression (\ref{ssb4}) with the interchange $P^+\leftrightarrow P^-$, and thus the figure \ref{polmassafter} is valid with this interchange.
In this case the vacuum energies are given by
\begin{eqnarray}
V_R(0,\pm w_0)=\frac{-3m^4_R}{2\lambda_R}\frac{P^+P^-}{(\gamma^2+1)^2},\qquad V_H(0,\pm w_0)= \frac{-3m^4_R}{2\lambda_R}\frac{(P^+)^2}{(\gamma^2+1)^2};
\label{vacenergySS}
\end{eqnarray}
 therefore, $V_R(0,\pm w_0)$ coincides exactly with the first expression in Eq. (\ref{vacenergy}); $ V_H(0,\pm w_0)$ can be obtained from the second expression in Eq. (\ref{vacenergy}) by the change $P^-\rightarrow P^+$, and hence corresponds to the curve on the right hand side in the figure \ref{vacenergytwo}, but mirrored on the  $``y"$ axis. 

As a generic feature of almost all of the considered models we can generate the electroweak/Planck hierarchy by fine-tuning the $\gamma$-parameter and associating electroweak energy scales to one of the fields involved in the formalism and Planck mass scales to the other one. Conversely, in the attempt to obtain a natural hierarchy by fixing the $\gamma$-parameter through the critical points of the $\gamma$-polynomials, the physically relevant mass ratio of the aforementioned fields renders small hierarchies with values between 1 and $10^4$ at most.

\section {Hyperbolic deformation of electrodynamics}  
\label{hed}

In \cite{1} the following action was proposed as the hyperbolic deformation of the scalar electrodynamics  that describes a charged scalar field coupled to $U(1)$ gauge fields, 
\begin{equation}
     {\cal L} = - \frac{1}{4} F^{2}_{\mu\nu} +  | (\partial_{\mu} - ieA_{\mu})\psi|^{2}-V(\psi,\overline{\psi}),
     \label{QED}
\end{equation}
where the coupling constant $e$ was considered by simplicity as real, and the potential $V(\psi,\overline{\psi})$ is that considered previously in (\ref{complexlag}); the action is invariant under
the local gauge transformations
\begin{equation}
     \psi \rightarrow e^{i\theta} e^{j\chi} \psi, \qquad A_{\mu} \rightarrow A_{\mu} + \frac{1}{e} (\partial_{\mu}\theta - ij\partial_{\mu}\chi), \quad F_{\mu\nu}= \partial_{\mu}A_{\nu}-\partial_{\nu}A_{\mu} \rightarrow F_{\mu\nu},     
               \label{hgt}
\end{equation}
where in general the arbitrary real functions depend on the background space-time coordinates, $\theta = \theta (x)$, $\chi =\chi (x)$; hence, the original $U(1)$-symmetry is enlarged to $U(1)\times SO(1,1)$. The expanding around the vacuum $\psi \rightarrow \psi + \psi_{0}$, and $A_{\mu} \rightarrow A_{\mu}+ A^0_{\mu}$, leads to the expression
\begin{eqnarray}
     {\cal L} (\psi + \psi_{0}, A+A_{0}) \!\! & = & \!\! -\frac{1}{4} F^{2}_{\mu\nu} + e^{2}|\psi_{0}|^{2} B^{2}_{\mu} + \partial_{\mu}\psi \cdot \partial^{\mu} \overline{\psi} + B^{\mu} \Big \{ie(\overline{\psi}_{0}\partial_{\mu}\psi - \psi_{0}\partial_{\mu}\overline{\psi}) + 2e |\psi_{0}|^{2}(\partial_{\mu}\theta - ij \partial_{\mu}\chi) \nonumber \\
     \!\! & & \!\! + A^{0}_{\mu}(\psi_{0}\overline{\psi} + \overline{\psi}_{0}\psi)\Big\} +i (\partial^{\mu}\theta -ij\partial^{\mu}\chi)(\overline{\psi}_{0}\partial_{\mu}\psi - \psi_{0}\partial_{\mu}\overline{\psi})+ieA^{\mu}_{0} (\overline{\psi}\partial_{\mu}\psi - \psi\partial_{\mu}\overline{\psi})\nonumber \\
     \!\! & & \!\! + |\psi_{0}|^{2}(\partial_{\mu}\theta - ij\partial_{\mu}\chi) (\partial^{\mu}\theta -ij\partial^{\mu}\chi) -V(\psi+ \psi_{0},\overline{\psi}+\overline{\psi_0})+ {\rm higher \ terms},
     \label{qed2}
\end{eqnarray}
where $A_{\mu}$ has been replaced by $A_{\mu} = B_{\mu} + \frac{1}{e} (\partial_{\mu}\theta - ij\partial_{\mu}\chi)$, and $F_{\mu\nu}$ is expressed now in terms of the new field $B_{\mu}$. 

 After eliminating the interaction terms of the form $B^{\mu} \cdot \partial_{\mu} (\varphi , \overline{\varphi}, \theta , \chi)$, with the identification \cite{1}
\begin{eqnarray}
\chi=\frac{2\gamma}{\gamma^2-1}\frac{w_0^2-v_0^2}{w_0^2+v_0^2}\theta,
\nonumber\\
\theta = \frac{(\gamma^4-1)(w_0^2+v_0^2)}{(\gamma^2-1)^2(w_0^2+v_0^2)^2+4\gamma^2(w_0^2-v_0^2)^2}(w_0v-v_0w),
\label{thetavw}
\end{eqnarray}
 the quadratic terms in the Lagrangian reduce to
\begin{eqnarray}
    {\cal L} (\psi +\psi_{0}, A+A_{0}) \!\! & = & \!\! -\frac{1}{4} F^{2}_{\mu\nu} + e^{2}|\psi_{0}|^{2} B_{\mu}B^{\mu} + 
      (\gamma^2-1)(\partial v^{2} + \partial w^{2} )+2k\gamma (\partial w^{2} - \partial v^{2} )\nonumber\\    
     \!\! & +& \!\!  (\gamma^2+1)^2\frac{(1-\gamma^2)(w^2_0+v^2_0)+2k\gamma(w^2_0-v^2_0)}{(\gamma^2-1)^2(w^2_0+v^2_0)^2+4\gamma^2(w^2_0-v^2_0)^2}\Big[w_{0}^2\partial v^{2}-v_0 w_{0}\partial^{\mu}v\partial_{\mu}w+v_{0}^2\partial w^{2}\Big] \nonumber \\
     \!\! & & \!\!-\Big[ -\frac{am^2}{2}\psi\overline{\psi} + \frac{\lambda}{4!}(\overline{\psi}^{2}_{0} \psi^{2} + \psi^{2}_{0} \overline{\psi}^{2} )\Big]+ {\rm higher \ terms};    
     \label{exploitt}
\end{eqnarray}
the interacting terms of the form $\partial^{\mu}v\cdot\partial_{\mu}w$, are proportional to $v_{0}\cdot w_{0}$, which will vanish at the end due to only one v.e.v., $v_{0}$ or $w_{0}$ will acquire a non-zero value.
This quadratic expression is fully general for  arbitrary vacuum fields $(v_{0}\neq 0,w_{0}\neq 0)$, and hence we have not made an identification between the interaction parameters $(\lambda_{R}, \lambda_{H})$, or mass parameters $(m^2_R, m^2_H)$. In \cite{1}, the case with ($v_{0}\neq 0, w_{0}=0)$, and the identification $(\lambda_{R}= \lambda_{H})$ was studied exhaustively; we complement that study by considering now the case considered in the section \ref{no1} in this paper, namely, ($v_{0}\neq 0, w_{0}=0)$, but with an identification of the mass parameters.

 \subsection{ ($v_{0}\neq 0, w_{0}=0)$, and $m^2_R=m^2_H$.}
\label{qedvacuum}  
Note that in the case of local rotations, the under-braced $vw$-terms in the expression (\ref{quacub1}), will correspond to terms of higher orders
than the quadratic ones; thus, the expressions (\ref{usualchoice}), $(\ref{diagonal2})$, and $(\ref{diagonal3})$ can not be used;  it is not possible to retain the mass hierarchy constructed previously for the local case.

\begin{figure}[H]
  \begin{center}
  \includegraphics[width=.6\textwidth]{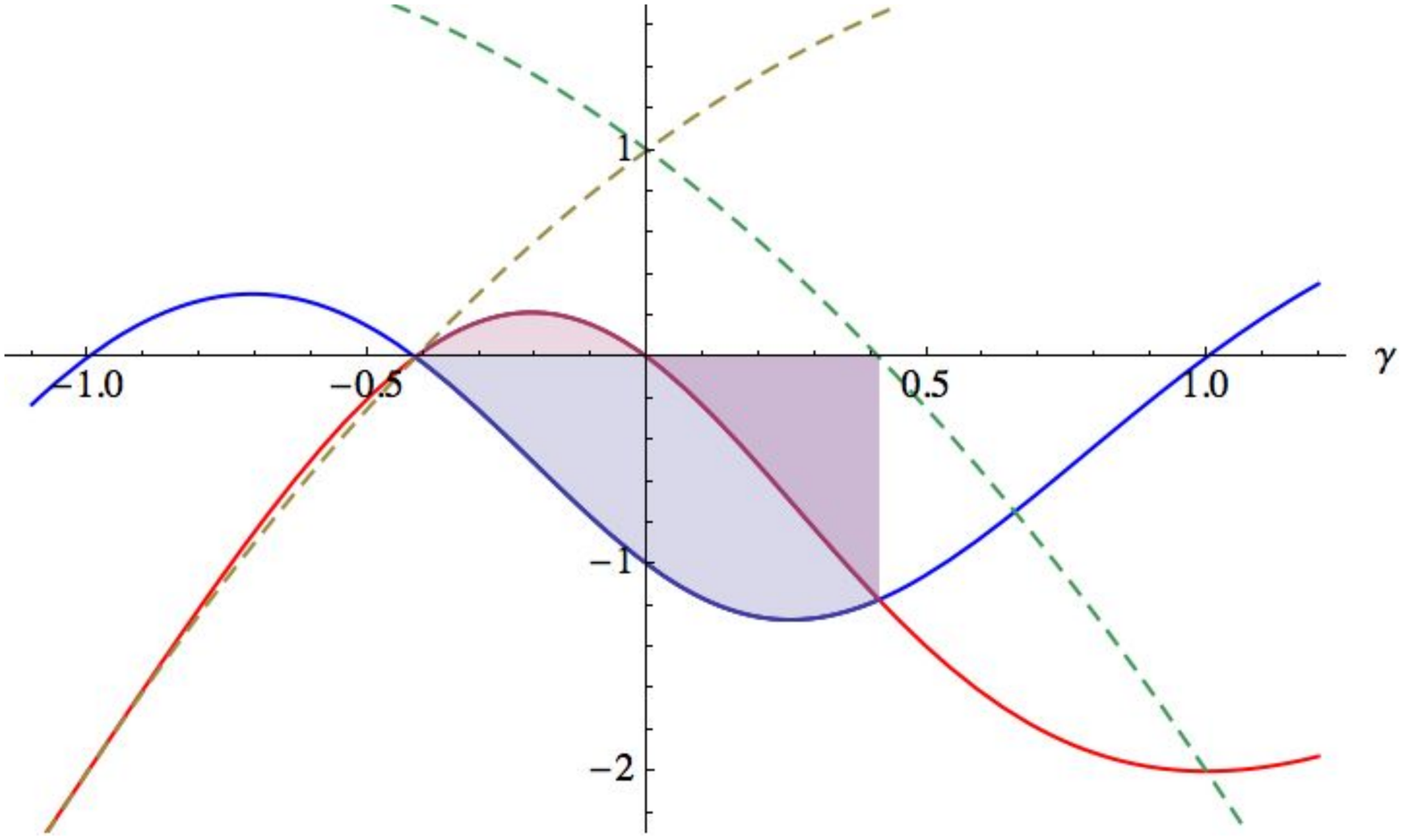}
\end{center}
\caption{ The blue curve represents $M^B_{R}(\gamma)$ , and the red curve $M^B_{H}(\gamma)$; the dashed curve represents the Hermitian mass for the field $v$ in Eq. (\ref{exploit}) (see figure \ref{polmassafter}); the case $\gamma=0$ reproduces the usual SSB of $U(1)$, with $M^B_{R}(0)=-1$, and $M^B_{H}(0)=0$.}
\label{massBB}
\end{figure}

\begin{figure}[H]
  \begin{center}
  \includegraphics[width=.6\textwidth]{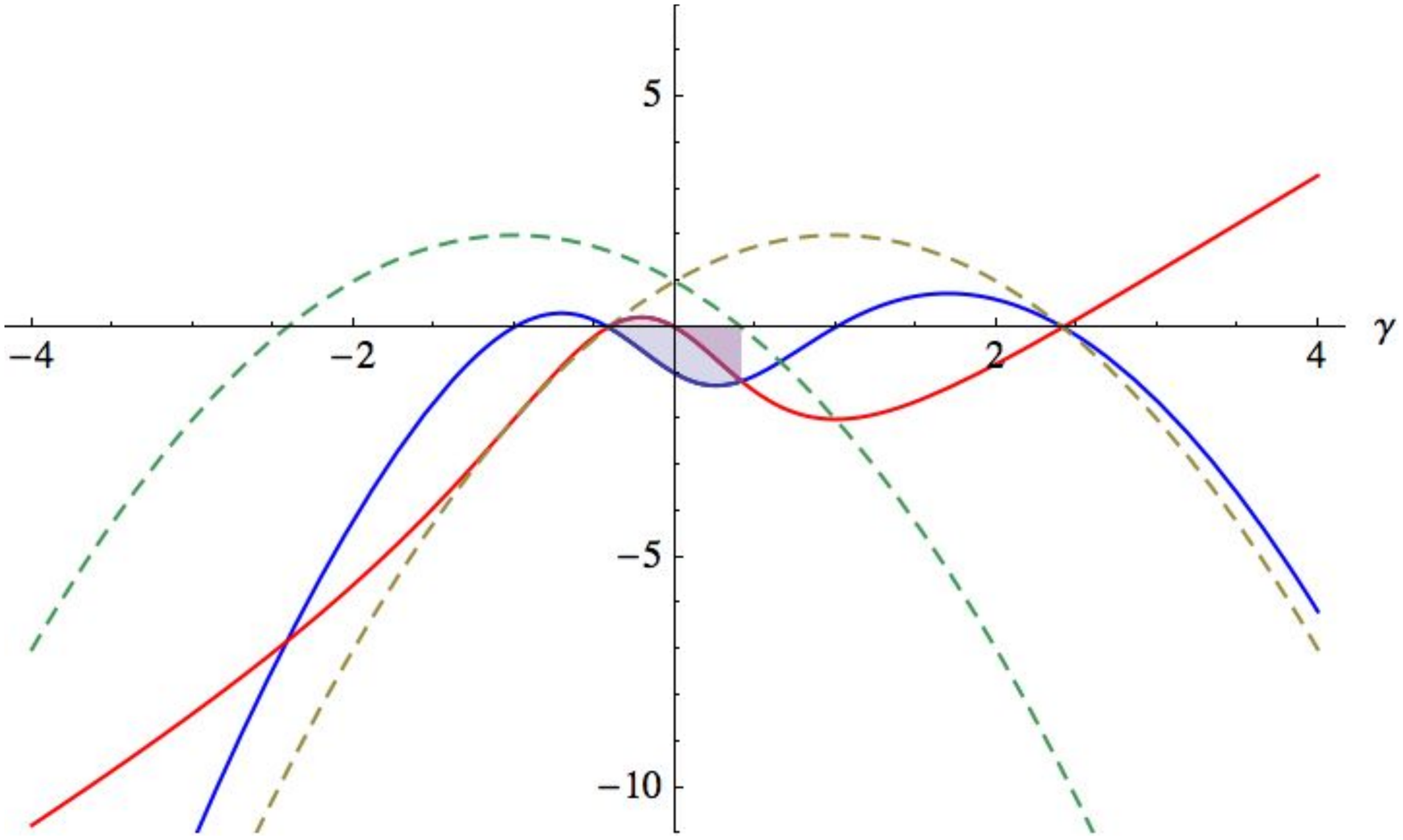}
\end{center}
\caption{ Far from the interval $(1-\sqrt{2},\sqrt{2}-1)$, $M^B_{R}(\gamma)$  behaves as an inverted parabola, $lim_{\gamma\rightarrow \pm \infty}M^B_{R}\rightarrow -\gamma^2$;  $M^B_{H}(\gamma)$ behaves as a line, $lim_{\gamma\rightarrow \pm \infty}M^B_{H}\rightarrow 2\gamma$.}
\label{mass4}
\end{figure}

The quadratic mass term (\ref{ssb4}) is valid for local rotations, by approaching the circular and hyperbolic parameters to first order \cite{1}; the Lagrangian (\ref{exploitt}) reduces to
\begin{eqnarray}
     {\cal L} (\psi +\psi_{0}, A+A_{0}) \!\! & = & \!\! -\frac{1}{4} F^{2}_{\mu\nu} + e^{2}|\psi_{0}|^{2} B_{\mu}B^{\mu} + 
      (\gamma^2-1-2k\gamma)\partial v^{2}+a(-P^+-kP^-) m^2_Rv^2 \nonumber \\  
       \!\! & + & \!\! {\rm higher \ terms};    
     \label{exploit}
\end{eqnarray}
since the kinetic and the mass terms  for the field $w$ have disappeared, it corresponds  to a Nambu-Goldstone field. 
The v.e.v. of the Higgs field $v$, given by the expression (\ref{positivequan1}),
 determines the (Hermitian) mass of the longitudinal mode of the field $B$,
 \begin{eqnarray}
  e^{2}|\psi_{0}|^{2}=e^2(\gamma^2-1-2k\gamma)v^2_{0}=\frac{6e^2}{a\lambda_R}\Big[ M^{B}_{R}(\gamma)+kM^{B}_{H}(\gamma)
  \Big]m^2_{_{R}}; \nonumber\\
  M^{B}_{R}(\gamma)\equiv \frac{1-\gamma^2}{\gamma^2+1}P^- , \quad M^{B}_{H}(\gamma)\equiv \frac{2\gamma}{\gamma^2+1}P^-; 
    \label{massesB}
   \end{eqnarray}
The Hermitian vector field $B$ with the form $B\equiv B_{R}+kB_{H}$, has acquired a Hermitian mass through Higgs mechanism;  the figure \ref{massBB} shows the behavior close to the interval $(1-\sqrt{2},\sqrt{2}-1)$, and the figure \ref{mass4} the global behavior. The polynomial masses have no asymptotes, as opposed to the case studied in \cite{1} with the restriction $\lambda_R=\lambda_H$; the asymptotes in that case are just at the limits of the interval $(1-\sqrt{2},\sqrt{2}-1)$. 

 The real mass $M^{B}_{R}$ vanishes at the two obvious roots  $\gamma^2=1$, that correspond to the purely hyperbolic limit for the theory \cite{1}; however,  these roots are out of the interval $(1-\sqrt{2},\sqrt{2}-1)$. Additionally it vanishes at the roots of $P^-$; in particular at the left limit $1-\sqrt{2}$, the $k$-mass $M^{B}_{H}$, and the $k$-mass of $v$ also vanish, and we shall have a nearly massless vectorial field, and a massive $v$ field, since its real mass survives at this limit. We must remember that strict masslessness is forbidden, since the Hessian of the zero-energy point for the potentials vanishes, and the profile of the potentials with a stable degenerated vacuum, and with an energy bounded from below  is lost. 
  
 At the right limit $\sqrt{2}-1$, the vectorial field is massive in both senses, and the values of its real and $k$-masses coincide; in this limit the Higgs field $v$ is only $k$-massive;   
 \begin{eqnarray}
  \frac{6e^2}{a\lambda_R}\Big[ M^{B}_{R}(\sqrt{2}-1)+kM^{B}_{H}(\sqrt{2}-1)
  \Big]m^2_{_{R}}= \frac{6e^2}{a\lambda_R}\Big[ 2(\sqrt{2}-2)+2(\sqrt{2}-2)k
  \Big]m^2_{_{R}},\nonumber\\
  -P^-(\sqrt{2}-1) = 4(\sqrt{2}-1);
    \label{higher}
    \end{eqnarray} 
with $2(\sqrt{2}-2)\approx -1.1716$, and  $ 4(\sqrt{2}-1)\approx 1.6569$; thus, there is no hierarchy at this point.

Furthermore, $ M^{B}_{R}$ has a local minimum at $\gamma\approx 0.25605$;
\begin{eqnarray}
 M^{B}_{R} (\gamma\approx 0.25605)\approx -1.2685, \quad  M^{B}_{H} (\gamma\approx 0.25605)\approx -0.6951; \nonumber \\
  -P^+(\gamma\approx 0.25605) \approx  0.4223, \quad   -P^-(\gamma\approx 0.25605) \approx 1.4465;
  \label{localmb}
 \end{eqnarray}
 therefore, the real mass of $B$ is of the same order than the $k$-mass of $v$, and similarly the $k$-mass of $B$
 is of the same order than the real mass of $v$, and a hierarchy of order $10$ is induced; these values for the masses of $v$ must be compared with those values in Eq. (\ref{vev}), calculated at the maximum value of v.e.v.  $v_0$, which is very close to the value of $\gamma$ considered in the above expressions. 
 Similarly, along the same lines, we can determine the hierarchy at the local maximum of the $M^B_{H}(\gamma)$ within the interval $(1-\sqrt{2},\sqrt{2}-1)$. 
However, all these critical points lie out of the restricted range (\ref{newran}), and thus there are no natural criteria for defining a hierarchy. However, there is another possibility with respect to the expression (\ref{massesB}), since the distribution of the polynomials into the effective coupling parameters is fully arbitrary.

 \subsection{Flow for the electromagnetic coupling}
 The expression (\ref{massesB}) can be rewritten as
  \begin{eqnarray}
  e^{2}|\psi_{0}|^{2}=\frac{6}{a\lambda_R}\frac{e^2}{\gamma^2+1}\Big[ M^{B}_{R}(\gamma)+kM^{B}_{H}(\gamma)
  \Big]m^2_{_{R}}; \nonumber\\
  M^{B}_{R}(\gamma)\equiv (1-\gamma^2)P^- , \quad M^{B}_{H}(\gamma)\equiv 2\gamma P^-; 
    \label{massesB2}
   \end{eqnarray}  
where we have re-defined the effective masses, and we have additionally an effective form for the electromagnetic charge; however, the new masses
will have essentially the same hierarchy. 
 \begin{figure}[H]
 \begin{center}
  \includegraphics[width=.5\textwidth]{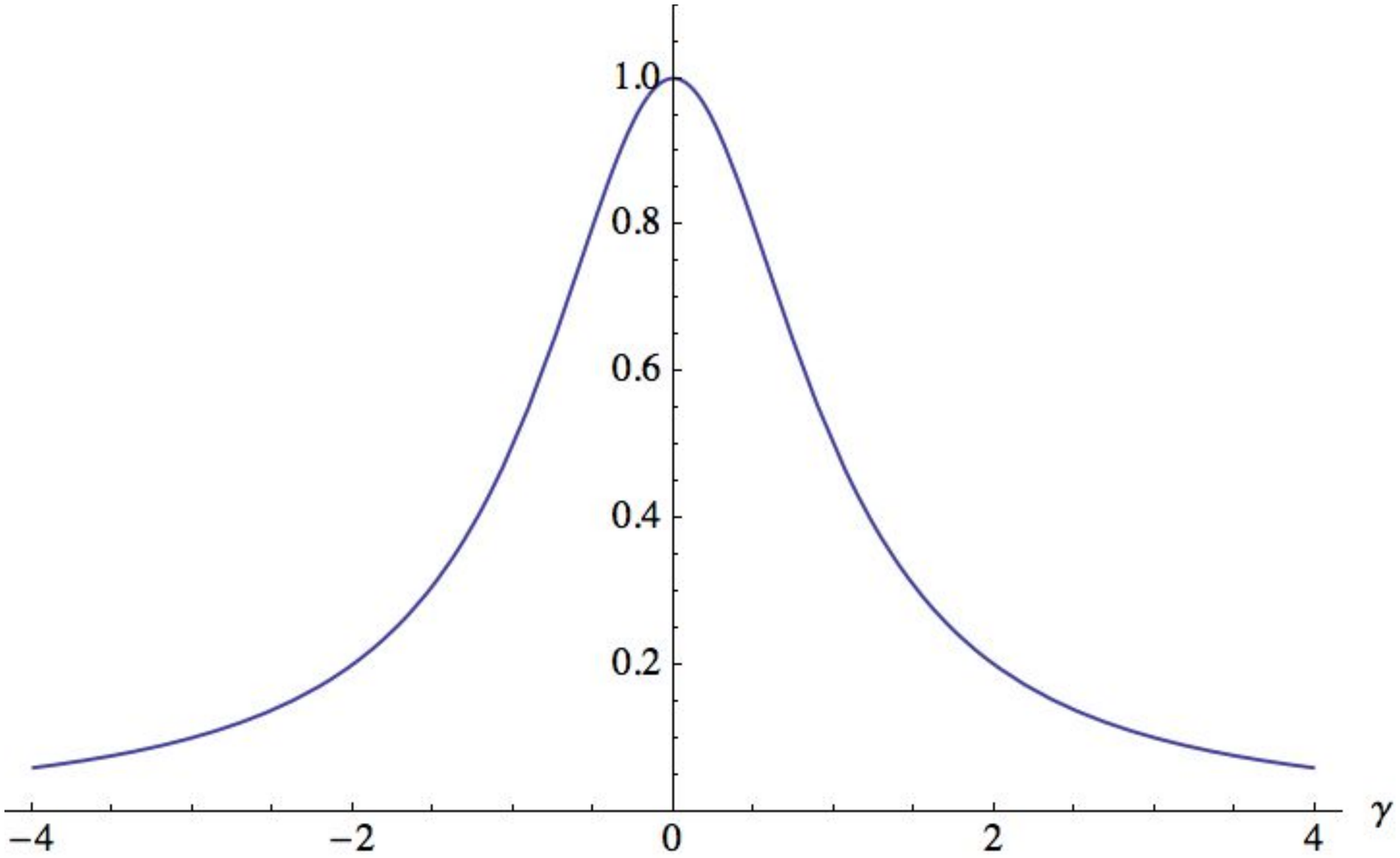}
  \end{center}
\caption{The polynomial effective charge $\frac{[e^2]}{(\gamma^2+1)}$.}
\label{barecharge}
\end{figure}
On the other hand, the $\gamma$-deformation effectively diminishes the charge, mimics a renormalization flow for the coupling $e$; for example, one can compare this expression with the normalized charge squared in QED, in some units system,
\begin{equation}
e^2_R=e^2\frac{1}{1+\frac{e^2}{12\pi^2}ln(M^2/m^2)},
\end{equation}
where $M$ is the energy scale, and $m$ the electron mass; the coincidence in the functional dependence suggests some relation between the $\gamma$-parameter and the renormalization scale; this issue will be the subject of forthcoming works.

The effective charge goes asymptotically to zero as $\gamma\rightarrow\pm\infty$; however, we have the restricted range (\ref{newran}), and hence we can determine the minimum value for the effective charge at the limits of the range,
\begin{equation}
\frac{1}{1 + (2 - \sqrt{6}+ \sqrt{5 - 2 \sqrt{6}})^2}\approx 0.982963,
\end{equation}
thus, the $\gamma$-effective value of the coupling constants can be comparable with those generated by quantum fluctuations.
  
\section{Concluding remarks} 
\label{cr}

In the hypercomplex formulation of gauge field theories, a compact gauge group is deformed through a $\gamma$-parameter that varies along a non-compact internal direction, transverse to the $U(1)$ compact one; in this manner, a non-compact gauge group is accommodated, and the invariant objects under the action of the full group, $U(1)\times SO(1,1)$, are necessarily Hermitian, generalizing the usual real invariant objects. 

This non-compact internal direction can effectively be interpreted as an internal extra dimension which controls the spontaneous symmetry breaking of the theory through a $\gamma$-deformation parameter that allows us to establish a mass hierarchy by assigning Panck mass scales for one field and electroweak mass scales for the other field of the formalism. Both fields are considered fundamental in this scheme, however, the solution to the hierarchy problem is achieved through a fine-tuning of the aforementioned $\gamma$-parameter as a quite generic feature of the model. 

On the other hand, when trying to get a natural hierarchy by fixing the $\gamma$-parameter through the critical points of the $\gamma$-polynomials, the relevant mass ratio of the above mentioned fields yields a small hierarchy with values that vary between unity and $10^4$.

We would like to remark as well that in order to simplify the analysis, in Section (\ref{gammano1}) we have made the assumption that 
$\lambda_R=\lambda_H$, which leads to a direct cancelation of such parameters in the Eq. (\ref{expvalue}), and thus the mass ratio is depending only on the $\gamma$-parameter. However, a more general case can be considered by defining the quotient $\lambda\equiv \frac{\lambda_H}{\lambda_R}$, and hence the physically relevant mass ratio reduces to 
\begin{equation}
\frac{m^2_H}{m^2_R}=\frac{(1-\gamma^2)\lambda-2\gamma}{(1-\gamma^2)-2\gamma\lambda};
\end{equation}
therefore the possible mass hierarchies are depending now on the choice, and possibly the fine tuning, of two free parameters $(\gamma,\lambda)$.
A similar generalization can be made in relation to the simplifications considered in Sections (\ref{no1}), (\ref{nogamma11}), and (\ref{hed}).
These generalized cases yield qualitative and quantitatively different scenarios for the hierarchy, and will be considered in future communications.\\

{\bf Acknowledgements:}
This work was supported by the Sistema Nacional de Investigadores (M\'exico), and the Vicerrectoria de Investigaci\'on y Estudios de Posgrado (BUAP). AHA also thanks PRODEP for partial financial support, and AEH acknowledges the support by CONACyT under Grant No. CB-2014-01/ 240781.
 The numerical analysis and graphics have been made using Mathematica.


\begin{thebibliography}{}
\setlength{\itemsep}{-.50em} \setlength{\itemsep}{-.50em}
\bibitem{emam1} M. H. Emam, {\it BPS one-branes in five-dimensions}, [hep-th/1108.339].
\bibitem{emam2} M. H. Emam, {\it Wrapped M5-branes leading to five dimensional 2-branes}, Phys. Rev. D {\bf 74}, 125004 (2006).
\bibitem{emam3} M. H. Emam, {\it Symplectic covariance of the N hypermultiplets}, Phys. Rev. D {\bf 79}, 085017 (2009).
\bibitem{emam4} M. H. Emam, {\it Five dimensional 2-branes from special Lagrangian wrapped M5-branes}, Phys. Rev. D {\bf 71}, 125020 (2005).
\bibitem{emam5} M. H. Emam, {\it Split-complex representation of the universal hypermultiplet
}, Phys. Rev. D {\bf 84}, 045016 (2011).
\bibitem{perry} G. W. Gibbons, M. B. Green, M. J. Perry, {\it Instantons and Seven-Branes in Type IIB Superstring Theory}, Phys. Lett. B {\bf 370}, (1996), 37 [hep-th/9511080]
\bibitem{ulrych} S. Ulrych,{\it  Considerations on the hyperbolic complex KleinÐGordon equation
},  J. Math. Phys. {\bf 51}, 063510 (2010).
\bibitem{kisil} V. V. Kisil, {\it Starting with the group $SL_{2}(R)$}, Notices of the AMS, Vol. {\bf 54}, number 11 (2007).
\bibitem{1} R. Cartas-Fuentevilla, and O. Meza-Aldama, {\it Spontaneous symmetry breaking, and strings defects in hypercomplex gauge field theories}, Eur. Phys. J. C,  {\bf 76}, 98, (2016) [hep-th/1506.04410]. 
\bibitem{vergara} C. A. Margalli, and J. D. Vergara, {\it Hidden symmetries in holomorphic models}, Phys. Lett. A. {\bf 379}, 2434 (2015).
\bibitem{wittensusy} E. Witten, {\it Mass hierarchies in supersymmetric theories}, Phys. Lett. B {\bf 105}, 267 (1981).
\bibitem{anto} I. Antoniadis, {\it  A possible new dimension at a few TeV }, Phys. Lett. B {\bf 246}, 377 (1990).
\bibitem{nima} N. Arkani-Hamed, and M. Schmaltz, {\it  Hierarchies without symmetries from extra dimensions}, Phys. Rev. D {\bf 61}, 033005 (2000).
\bibitem{antonima} I. Antoniadis, N. Arkani-Hamed, {\it New dimensions at a millimiter to a Fermi and superstrings at a tev}, Phys. Lett. B {\bf 436}, 24 (1998).
\bibitem{rs} L. Randall, and R. Sundrum, {\it  Large mass hierarchy from a small extra dimension}, Phys. Rev. Lett. {\bf 83},  (1999), 3370.

\bibitem{bhnkq} N. Barbosa-Cendejas, A. Herrera-Aguilar, K. Kanakoglou, U. Nucamendi and I. Quiros,
    {\it Mass hierarchy and mass gap on thick branes with Poincar\'{e} symmetry}, Gen. Rel. Grav. {\bf 46}, 1631 (2014.)

\bibitem{dewolfe} O. DeWolfe, D.Z. Freedman, S.S. Gubser and A. Karch, {\it Modeling the fifth dimension with scalars and gravity},
    Phys. Rev. D {\bf 62}, 046008 (2000).

\end{thebibliography}
\end{document}